\documentclass[preprint,11pt]{elsarticle}
\usepackage[top=1in, bottom=1in, left=1in, right=1in]{geometry}

\usepackage{graphicx}
\usepackage{amssymb}

\usepackage[colorlinks=true,bookmarksopen,pdfsubject={algorithms},linkcolor={blue}, anchorcolor={black}, citecolor={blue}, filecolor={magenta}, menucolor={black}, pagecolor={red},backref=none,urlcolor={blue}]{hyperref}

\usepackage{lineno}
\usepackage{comment}
\usepackage{amsmath}
\usepackage{tabu}
\usepackage{setspace}
\usepackage{bm}
\usepackage{appendix}
\usepackage[utf8]{inputenc} 
\usepackage[T1]{fontenc}    
\usepackage{url}            
\usepackage{booktabs}       
\usepackage{amsfonts}       
\usepackage{nicefrac}       
\usepackage{microtype}      
\usepackage{microtype}
\usepackage{mathrsfs}
\usepackage{graphicx}
\usepackage{subfigure}
\usepackage{amssymb}
\usepackage{booktabs} 
\usepackage{bm}

\usepackage{bm}
\usepackage{amsmath}
\usepackage{amssymb}
\usepackage{setspace}
\usepackage{bigints}
\usepackage{array}
\usepackage{multirow}
\usepackage{colortbl}
\usepackage{graphicx}
\usepackage{subfigure}
\usepackage[table,xcdraw]{xcolor}
\usepackage{booktabs}
\usepackage{tabu}
\usepackage{bbm}
\usepackage{float}
\usepackage[singlelinecheck=false]{caption}
\usepackage[utf8]{inputenc}
\usepackage[english]{babel}
\usepackage{bookmark}
\usepackage{mathrsfs}
\usepackage{color}
\usepackage{mathtools} 
\usepackage{amssymb}
\usepackage{xfrac}
\usepackage{lscape}
\usepackage{pdflscape}
\usepackage{algorithm}
\usepackage{algpseudocode}
\usepackage{xspace}
\usepackage{booktabs}

\usepackage{natbib}

\journal{Materials \& Design}

\begin{document}
\begin{spacing}{1.15}
\begin{frontmatter}

\title{A directional Gaussian smoothing optimization method for computational inverse design in nanophotonics  \footnote{\footnotesize{This manuscript has been co-authored by UT-Battelle, LLC, under contract DE-AC05-00OR22725 with the US Department of Energy (DOE). The US government retains and the publisher, by accepting the article for publication, acknowledges that the US government retains a nonexclusive, paid-up, irrevocable, worldwide license to publish or reproduce the published form of this manuscript, or allow others to do so, for US government purposes. DOE will provide public access to these results of federally sponsored research in accordance with the DOE Public Access Plan (http://energy.gov/downloads/doe-public-access-plan).}}}

\author[Second]{Jiaxin Zhang\corref{cor1}}
\ead{zhangj@ornl.gov}

\author[Fourth,Fivth]{Sirui Bi\corref{cor1}}
\author[Second]{Guannan Zhang}

\cortext[cor1]{These two authors contributed equally to this work.}

\address[Second]{Computer Science and Mathematics Division, Oak Ridge National Laboratory, Oak Ridge, TN 37830, USA}
\address[Fourth]{Computational Science and Engineering Division, Oak Ridge National Laboratory, Oak Ridge, TN 37830, USA}
\address[Fivth]{Department of Civil and Systems Engineering, Johns Hopkins University, Baltimore, MD 21210, USA}


\begin{abstract}

Local-gradient-based optimization approaches lack nonlocal exploration ability required for escaping from local minima in non-convex landscapes. A directional Gaussian smoothing (DGS) approach was recently proposed by the authors (Zhang et al., 2020) and used to define a truly nonlocal gradient, referred to as the DGS gradient, in order to enable nonlocal exploration in high-dimensional black-box optimization. Promising results show that replacing the traditional local gradient with the nonlocal DGS gradient can significantly improve the performance of gradient-based methods in optimizing highly multi-modal loss functions. However, the current DGS method is designed for unbounded and unconstrained optimization problems, making it inapplicable to real-world engineering design optimization problems where the tuning parameters are often bounded and the loss function is usually constrained by physical processes. In this work, we propose to extend the DGS approach to the constrained inverse design framework in order to find a better design. The proposed framework has its advantages in portability and flexibility to naturally incorporate the parameterization, physics simulation, and objective formulation together to build up an effective inverse design workflow. A series of adaptive strategies for smoothing radius and learning rate updating are developed to improve the computational efficiency and robustness. To enable a clear binarized design, a dynamic growth mechanism is imposed on the projection strength in parameterization. Our methodology is demonstrated by an example of designing a nanoscale wavelength demultiplexer and shows superior performance compared to the state-of-the-art approaches. By incorporating volume constraints, the optimized design achieves an equivalently high performance but significantly reduces the amount of material usage.

\end{abstract}


\end{frontmatter}

\section{Introduction}
\label{S:1}
Photonic devices have been applied in a wide range of applications, including photonic integrated circuits \cite{coldren2012diode}, optical interconnects and sensors \cite{miller2010optical,lin1997porous}, augmented reality \cite{kress2013review} and quantum computing \cite{kok2007linear}. As a growing number of applications in nanophotonic devices, photonic design is becoming increasingly demanding and challenging to optimize the device performance \cite{polman2012photonic, Molesky2018,wu2018design, yang2020inverse}. Classical design approaches based on analytical theory and intuition, however, are limited in small design space and relatively simple parameter tuning by hand. Capitalizing on the increased degrees of freedom in design space, nanophotonic devices have been designed with novel functionalities, high performance, efficiencies, and robustness that have been proven difficult to implement in traditional intuition-based methods \cite{Piggott2015}. 

There recently have been significant interests in using {\em computational inverse design} approaches to explore the full design space of novel photonic devices with a broad variety of applications \cite{Su2020, Peurifoy2018, jensen2011topology, Angeris2019, phan2019high, frandsen2014topology, Hughes2018, yan2019flexible, frellsen2016topology, Su2018,jiang2019simulator}. Much of this progress is made by the gradient-based algorithm, which is a promising method to efficiently search the enormous degrees of freedom in high-dimensional design space. The gradient-based optimization typically relies on adjoint method \cite{jensen2004systematic,borel2004topology}, which is a technique that enables the local gradient of an objective function and constraints to be calculated with respect to arbitrarily large design variables using forward physical simulation such as electromagnetic simulations. To this end, several recent studies further use automatic differentiation and backpropagation tools that are beneficial from machine learning research, to efficiently evaluate the local gradient by reducing the number of simulations \cite{Su2020, Hughes2018,hughes2019forward,minkov2020inverse}. These approaches make a feasible gradient-based design of photonic structures, particularly nanophotonic devices, with better efficiency and smaller footprints than traditional devices. {\color{black} However, these approaches basically depend on an efficient estimation or analytical derivation of the gradient (or called sensitivity analysis). Typically, several assumptions are often made for simplicity to utilize gradient-based optimization in the context of electromagnetism. For example, electromagnetism is modeled using Maxwell's equations assuming statics, linear, homogeneous, and isotropic materials as well as time-harmonic behavior of the field and transverse electric and magnetic problems with material invariance in the polarization direction \cite{christiansen2020tutorial,christiansen2020code}. All of these assumptions lead to a large challenge in real-world photonic design with dynamic, nonlinear, and dispersive material properties in complex multiphysics conditions. In addition, for complicated non-convex objective functions and constraints, gradient estimation relies on the adjoint method may be either not easily accessible or unreliable. Sometimes, additional efforts are required to derive the sensitivity analysis if unusual objectives or constraints are incorporated into the optimization formulation even though the existing gradient-based scheme has been used in device design. 
}

{\color{black} Another important challenge is that, up to now, most of the studies use local gradient-based approaches for inverse design so that the optimized devices converge to a \textit{local minimum} with respect to the design parameters.} In many electromagnetic design problems, their landscapes have been proven to be highly nonlinear and non-convex such that many possible local minima exist \cite{Molesky2018, Su2020}. These local minima depend on the initialization and vary largely as the {\color{black}initial guess} change. These challenges in gradient-based approaches have attracted much attention \cite{Su2020,jiao2006systematic, elesin2012design, campbell2019review}. In practice, one common way is to run an optimization several times with different {\color{black}initial guesses} that provide a rough estimate of the device performance. However, this approach has limitations in maximizing device performance and computational efficiency, and meanwhile, it possibly gives rise to a large variation of design performance. {\color{black} Alternatively, several evolutionary algorithms, including Generic Algorithm (GA), Particle Swarm Optimization (PSO) and Simulated Annealing (SA) are used to explore the global minima but finding the optimal solution to complex high-dimensional, multimodal problems often converges very slowly and requires very expensive fitness function evaluations \cite{schneider2019benchmarking}. Some recent studies optimize the photonic device performance using derivative-free methods, including Bayesian optimization (BO) \cite{sakurai2019ultranarrow} and differential evolution (DE) algorithms \cite{schneider2019benchmarking}, but these methods have limitations in scaling to high-dimensional problems \cite{snoek2012practical,price2006differential} in photonic device design. Therefore, it is necessary to develop an efficient optimization algorithm that can also escape from local minima in non-convex, high-dimensional landscape, to overcome the challenges in the local gradient-based approaches, evolutionary algorithms and derivative-free global optimization methods. 

}

In principle, it is feasible to optimize the photonic devices directly by changing the value of permittivity distribution at every point. However, it is more critical to impose fabrication constraints into optimization workflow because a fundamental challenge in nanophotonic device design is that arbitrary permittivity distribution, such as very tiny feature and grey-scale value, can not be fabricated in practice \cite{borel2004topology, vercruysse2019analytical,piggott2017fabrication}. The difficulty is often addressed by choosing an appropriate parameterization via a series of transformations that are simply an operation that affects the state of optimization. The use of transformation allows different parameterization and optimization stages to be easily swapped in and out for one another. This requires the optimizer has the capability of naturally integrating transformation into the design process. Another common constraint in optimizing material layout is the material usage (or volume fraction) in device design. In other words, it is desirable to use fewer materials but able to achieve performance that is as good as the target. To the author’s knowledge, relatively few studies have accounted for the problem of inverse design with volume constraint in nanophotonic design. One possible solution is to add a penalty term into objective function and thus convert the unconstrained optimization to constrained optimization through the augmented Lagrangian formalism. However, the penalty coefficient is very sensitive and typically difficult to control in practical implementation. {\color{black} Although recent advances in stochastic optimization algorithms, e.g., SGD, Adam, and RMSProp, have attracted much attention and widely used in machine learning training, it is a non-trivial task to incorporate multiple equality or inequality constraints into these stochastic methods that mainly focus on unconstrained optimization problems.}

{\color{black} 
To address these challenges, we propose a nonlocal inverse design workflow by incorporating the nonlocal gradient that was recently developed in in \cite{2020arXiv200203001Z}. The nonlocal gradient was defined by directional Gaussian smoothing, thus it is referred to as the DGS gradient hereinafter. The DGS gradient conducts 1D nonlocal explorations along with d orthogonal directions and each of which defines a nonlocal directional derivative as a 1D integral. The d directional derivatives are assembled to the DGS gradient. The Gauss-Hermite (GH) quadrature is used to approximate the 1D integrals (i.e., the directional derivatives) to achieve higher accuracy than Monte Carlo (MC) sampling. We improved the existing DGS approach from two perspectives in the context of inverse design. First, we established a workflow in which the DGS gradient can be combined with a variety of constraints, e.g., fabrication constraints and materials usage constraint in the practical material design. Second, we developed a series of adaptive strategies for the smoothing radius and the learning rate in order to improve computational efficiency and robustness. Compared to the local gradient, the directional smoothing allows for a large smoothing radius to capture the global structure of loss landscapes and thus provide a strong nonlocal exploration capability for escaping from local minima in non-convex landscapes. Furthermore, our workflow does not rely on the sensitivity analysis with multiple assumptions, so that it has wider feasibility to nonlinear, dynamic, and non-isotropic materials under complex physical conditions. In the meantime, our method having the benefits of gradient-based optimization can be easily scaled to high-dimensional design spaces, which alleviates the challenges in derivative-free global optimization, such as Bayesian optimization.

}



The paper is structured as follows. Section 2 provides a brief mathematical formulation and overview of inverse design. Section 3 presents the DGS gradient operator in principle and explain how DGS gradient optimization can overcome the challenges and difficulties in the local gradient optimization approaches. In Section 4, we show an example of designing wavelength demultiplexers, explain the implementation of the nonlocal optimization method using the DGS gradient in detail, and demonstrate the strength and advantages via the discussion and comparison. {\color{black}After that, we provide a discussion section to address the implementation concern and current limitations of the proposed method.} Finally, we provide a brief conclusion and discussion of future work.  

\section{Mathematical formulation of inverse design}
\label{sec:2}

This section provides a brief overview of the mathematical foundations behind the inverse design in photonic devices. Although the exact optimization problem may vary from case to case, the photonic design generally shares a similar set of features and steps, which include formulating an optimization problem, incorporating fabrication constraints and parameterization, and solving the inverse problem by optimization. 

\subsection{Optimization problem formulation}
A general electromagnetic design problem can be cast into the following optimization formulation:
{\color{black}
\begin{equation}
\begin{aligned}
    \min_{\mathbf{x}}& \quad f(\mathbf{E}_1,...,\mathbf{E}_n, \bm{\epsilon}_1,...,\bm{\epsilon}_n, \mathbf{x}) \\
    \textup{subject~to} & \quad  g_j(\mathbf{x}) = 0, \quad  j = 1,...,m \\
    & \quad h_k(\mathbf{x}) \le 0, \quad k = 1,...,l \label{eq:opt}
\end{aligned}
\end{equation}
}
where $\mathbf{E}_i$ is the electric field corresponding to the permittivity distribution $\bm{\epsilon}_i$, which depends on a parameterization vector $\mathbf{x} \in \mathbb{R}$ {\color{black} which is the computational design domain}, and $f$ is the objective function that defines the target of the optimization. A typical objective is to maximize the transmission, which is equivalent to minimize the negative 
\begin{equation}
    f_{obj}(\mathbf{x}) = -|\mathbf{c}^{\dagger } \mathbf{E}(\bm{\epsilon}(\mathbf{x}))|^2 \label{eq:obj1}
\end{equation}
where $\mathbf{c}^{\dagger } \mathbf{E}$ means the overlap integrals to compute the model coupling efficiency of the electric field $\mathbf{E}$ with the target mode at the output. {\color{black} $h_k(\mathbf{x})$ and $g_j(\mathbf{x})$ in Eq.~\eqref{eq:opt} are inequality and equality constraints on $\mathbf{x}$, particularly fabrication and volume of materials constraints, and the index $k$ and $j$ mean the number of inequality and equality constraints respectively. } {\color{black} For the optimization problem in Eq.~\eqref{eq:opt}, the electric fields $\mathbf{E}_i$ generated by the input permittivity  distribution $\bm{\epsilon}(\mathbf{x})$ should satisfy the Maxwell's equations in the frequency domain,
\begin{equation}
    \nabla \times \frac{1}{\mu}\nabla  \times \mathbf{E}_i - \omega_i^2 \bm{\epsilon} (\mathbf{x}) \mathbf{E}_i = -i \omega_i \mathbf{J}_i
    \label{eq:maxwell}
\end{equation}
where $i=1,...,n$ is the input modes, $\omega_i$ is the angular frequency, $\mu$ is the magnetic permeability of free space and $\mathbf{J}_i$ is the input source which injects the current mode into the input waveguide.} Eq.~\eqref{eq:maxwell} is often solved by electromagnetic simulation using the finite-difference frequency-domain (FDFD) method \cite{christ1987three} or finite-difference time-domain (FDTD) method \cite{sullivan2013electromagnetic}. {\color{black} Typically, the perfectly matched layer (PML) boundary condition as an artificial absorbing layer for wave equations, is used to truncate computational regions in numerical methods to simulate problems with open boundaries \cite{berenger1994perfectly}. In short, the computational inverse design problem can be addressed by solving the optimization problem in Eq.~\eqref{eq:opt}, which is to find optimal $\mathbf{x}$ to minimize the objective function, defined by Eq.~\eqref{eq:obj1}, subject to Maxwell's equations in Eq.~\eqref{eq:maxwell}, and fabrication constraints $g_j(\mathbf{x})$ and $h_k(\mathbf{x})$. }

\subsection{Parameterization and constraints in optimization}
Solving the optimization problem defined in Eq.~\eqref{eq:opt} led to continuously varying features of $\bm{\epsilon} (\mathbf{x})$, which is difficult for fabricating devices in practice. This is because the fabricated devices are typically composed of distinct materials so the permittivity can only take on certain discrete values and must keep the same along the vertical direction in fabrication with top-down lithography. Minimum feature size is another essential fabrication constraint. It is therefore critical to describe the permittivity distribution through a {\em parameterization} that addresses the fabrication challenges in device design \cite{Su2020}.

Parameterization basically consists of two key components: {\em projection} operator and {\em filtering} operator. Projection operator aims to convert the continuous features to a binary feature that better captures a clear ``0-1'' design, where ``0'' represents a {\em background} material and ``1'' represents a {\em foreground} material in permittivity distribution. This can be achieved by defining an operator through the equation 
\begin{equation}
  \bm{\epsilon} (\mathbf{x}) =  \bm{\epsilon}_b (\mathbf{x}) + \mathbb{H}(\mathbf{\varphi} (\mathbf{x})) \label{eq:fab}
\end{equation}
where $\bm{\epsilon}_b (\mathbf{x})$ is a permittivity background (constant) and $\mathbf{\varphi}(\mathbf{x})$ is a 2D slice of the permittivity distribution and ranges from 0 to 1. A possible projection operator $\mathbb{H}$ is using nonlinear penalty methods \cite{jensen2004systematic, jensen2011topology}. Filtering operator is often used to eliminate very tiny features and avoid to the formation of checker-board pattern in material layout \cite{sigmund1998numerical}. For example, level set methods \cite{Piggott2017,Vercruysse2019} construct a fabrication constraint penalty function for geometry representation of the devices. Using an appropriate parameterization, the fabrication constraints $h_k(\mathbf{x})$ can be imposed and naturally perform a binary device design. 

Another common constraint in optimizing material layout is the volume fraction of material usage, which is defined by 
{\color{black}
\begin{equation}
    h_1(\mathbf{x}) = V(\mathbf{x})/V_0 - \gamma \le 0 \label{eq:vol}
\end{equation}
}
where $V$ and $V_0$ are the expected material volume and design domain volume respectively, and $\gamma$ is the specific volume fraction. A simple solution of incorporating volume constraints into optimization is to add penalty terms into objective function and thus convert to unconstrained optimization so that several algorithms, for example, gradient descent, Adam, etc. can be used. However, the penalty coefficient, in fact, is very sensitive and difficult to determine in practical implementation. 

\subsection{Solving the inverse design problem}
The inverse design problem can be defined to find the best permittivity distribution $\bm{\epsilon}$ and the corresponding electric field $\mathbf{E}_i$ to maximize the device performance described by the objective function in Eq.~\eqref{eq:opt} and simultaneously satisfy the physics constraints in Eq.~\eqref{eq:maxwell}, fabrication constraints in Eq.~\eqref{eq:fab} and material volume constraints in Eq.~\eqref{eq:vol}. 
It is a challenging task to solve this kind of constrained optimization problem that involves large-scale, high-dimensional design degrees of freedom and a highly non-convex and non-linear landscape \cite{Su2020, zhang2019learning}. Many recent efforts have been made to develop gradient-based optimization techniques for addressing the challenges \cite{Su2020, hughes2018adjoint}. 

To perform gradient-based optimization, the gradient $df_{obj}/d\mathbf{x}$ is required. Note that $\mathbf{E}_i$ and $\bm{\epsilon}_i$ can be complex-valued, {\color{black}where $i$ means the $i$-th mode.} Suppose the objective function $f$ is real-valued, the gradient can be computed by 
\begin{equation}
    \frac{df_{obj}}{d \mathbf{x}} = \frac{\partial f_{obj}}{\partial \mathbf{x}} + 2\mathcal{R} \left[\sum_i \left(\frac{\partial f_{obj}}{\partial \mathbf{E}_i}   \frac{d \mathbf{E}_i}{d \mathbf{x}} + \frac{\partial f_{obj}}{\partial \bm{\epsilon}_i}   \frac{d \bm{\epsilon}_i}{d \mathbf{x}} \right) \right]
\end{equation}
where $\mathcal{R}[\cdot]$ denotes taking the real part. The derivative terms ${\partial f_{obj}}/{\partial \mathbf{E}_i}$, ${\partial f_{obj}}/{\partial \bm{\epsilon}_i}$ and ${d \bm{\epsilon}_i}/{d \mathbf{x}}$ depend on the form of the objective function but ${d \mathbf{E}_i}/{d \mathbf{x}}$ is always required in electromagnetic simulation. It is therefore necessary to derive the gradient for FDFD or other simulation methods. {\color{black} Given the FDFD equation in Eq.~\eqref{eq:maxwell}, we can rewrite it by 
\begin{equation}
    ({\Omega} - \omega^2 \textup{diag}(\bm{\epsilon})) \mathbf{E} = -i \omega \mathbf{J}
\end{equation}
where $\Omega$ is the {\em discretized} version of the $\nabla \times {\mu}^{-1}\nabla  \times$ operator. Differentiating by through by $\bm{\epsilon}$, we have 
\begin{equation}
    {\Omega} \frac{d \mathbf{E}}{d \bm{\epsilon}} - \left[ \frac{d \mathbf{E}}{d \bm{\epsilon}} \omega^2 \textup{diag}(\bm{\epsilon}) + \omega^2\textup{diag}(\mathbf{E})  \right] = 0. \label{eq:diff_max1}
\end{equation}
If we rearrange the Eq.~\eqref{eq:diff_max1}, we have
\begin{equation}
    ({\Omega} - \omega^2 \textup{diag}(\bm{\epsilon})) \frac{d \mathbf{E}}{d \bm{\epsilon}} =  \omega^2 \textup{diag}(\mathbf{E})
     \label{eq:diff_max2}
\end{equation}
and the simulation gradient ${d\mathbf{E}_i}/{d \mathbf{x}}$ is therefore derived by
\begin{equation}
    \frac{d \mathbf{E}_i}{d \mathbf{x}}  = \frac{d \mathbf{E}_i}{d \bm{\epsilon}_i} \frac{d \bm{\epsilon}_i}{d \mathbf{x}} =  ({\Omega} - \omega_i^2 \textup{diag}(\bm{\epsilon}_i))^{-1} \omega_i^2 \textup{diag}(\mathbf{E}_i) \frac{d \bm{\epsilon}_i}{d \mathbf{x}}. \label{eq:grad}
\end{equation}
}
Note that, the computing in Eq.~\eqref{eq:grad} is often computationally intensive because it requires the same number of electromagnetic simulations as the number of design degrees of freedom. Thanks to the development of automatic differentiation techniques in machine learning \cite{baydin2017automatic,paszke2017automatic}, efficient implementation relies on automatic differentiation and backpropagation are introduced to reduce the computational cost \cite{Su2020,hughes2018adjoint, hughes2019forward}.




\section{The DGS gradient for nonlocal optimization}
In this section, we describe the DGS gradient operator that was developed in our previous work \cite{2020arXiv200203001Z}. To better explain the direction Gaussian smoothing strategy, we briefly recall the standard Gaussian smoothing \cite{NesterovSpokoiny15} for estimating local gradients. Specifically, it starts by defining a smoothed loss function
\[
       F_{\sigma}(\mathbf{x}) = \mathbb{E}_{\bm u \sim \mathcal{N}(0, \mathbf{I}_d)} \left[F(\mathbf{x} + \sigma \bm u) \right], 
\]
where $\mathcal{N}(0, \mathbf{I}_d)$ is the $d$-dimensional standard Gaussian distribution, and $\sigma > 0$ is the smoothing radius. $F_{\sigma}(\mathbf{x})$ inherits many characteristics from $F(\mathbf{x})$, e.g., convexity, the Lipschitz constant.
Then, the gradient $\nabla F_\sigma(\mathbf{x})$ can be represented as an expectation and estimated by drawing $M$ random samples $\{\bm u_m\}_{m=1}^M$
from $\mathcal{N}(0,\mathbf{I}_d)$, i.e., 
\begin{equation}\label{e40}
    \nabla F_{\sigma}(\mathbf{x}) = 
    \frac{1}{\sigma}\mathbb{E}_{\bm u \sim \mathcal{N}(0, \mathbf{I}_d)} \left[F(\mathbf{x} + \sigma \bm u)\, \bm u\right] \approx \frac{1}{M\sigma}\sum_{m=1}^M F(\mathbf{x} + \sigma \bm u_m)\bm u_m.
\end{equation}
%
The Monte Carlo (MC) estimator in Eq.~\eqref{e40} is substituted into any gradient-based algorithm to update the state $\mathbf{x}$. {\em The major drawback is that the error of the MC estimator in Eq.~\eqref{e40} is on the order of $\varepsilon \sim \mathcal{O}(d \sigma /\sqrt{M})$.} When the dimension $d$ is large (e.g., on the order of thousands) and the computing budget (the upper bound of $M$) is given, practitioners often have to sacrifice 
a nonlocal smoothing effect (with a relatively big $\sigma$) that helps skipping local mimina to achieve a required accuracy. In other words, Eq.~\eqref{e40} is mostly used in the local regime with a small value for $\sigma$.

%

\subsection{The nonlocal DGS gradient operator}\label{sec:grad}
The DGS gradient was developed to alleviate the above challenge with the standard Gaussian smoothing. The key idea behind the DGS gradient is to conduct 1D nonlocal explorations along $d$ orthogonal directions in $\mathbb{R}^d$, each of which defines a nonlocal directional derivative as a 1D integral. The Gauss-Hermite quadrature, instead of MC sampling, is used to estimate the $d$ 1D integrals to achieve high accuracy.

Specifically, we first define a 1D cross section of $F(\mathbf{x})$ as
\begin{equation*}
G(y \,| \,{\mathbf{x}, \bm \xi}) = F(\mathbf{x} + y\, \bm \xi), \;\; y \in \mathbb{R},
\end{equation*}
where $\mathbf{x}$ is the current state of $F(\mathbf{x})$ and $\bm \xi$ is a unit vector in $\mathbb{R}^d$. The Gaussian smoothing of $G(y)$, denoted by $G_\sigma(y)$, is defined by
\begin{equation}
\label{eq10}
    G_{\sigma}(y \,| \,{\mathbf{x}, \bm \xi}) :=  \frac{1}{\sqrt{2\pi}} \int_{\mathbb{R}} G(y + \sigma v\, |\, \mathbf{x}, \bm \xi)\, {\rm e}^{-\frac{v^2}{2}}\, dv
     = \mathbb{E}_{v \sim \mathcal{N}(0, 1)} \left[G(y + \sigma v\, |\, \mathbf{x}, \bm \xi) \right].
\end{equation}
This is also the Gaussian smoothing of $F(\mathbf{x})$ along the direction $\bm \xi$ in the neighbourhood of $\mathbf{x}$. 
The derivative of $G_{\sigma}(y|\mathbf{x},\bm \xi)$ at $y = 0$ can be represented by a 1D integral
\begin{equation}\label{e4}
    \mathscr{D}[G_{\sigma}(0 \,|\, \mathbf{x}, \bm \xi)] 
     = \frac{1}{\sigma}\,\mathbb{E}_{v \sim \mathcal{N}(0,1)} \left[G(\sigma v \, | \, \mathbf{x}, \bm \xi)\, v\right],
\end{equation}
where $\mathscr{D}[\cdot]$ denotes the differential operator. We emphasize that Eq.~\eqref{e4} is fundamentally different from the directional derivative of $F_\sigma(\mathbf{x})$, because $G_{\sigma}(0 \,|\, \mathbf{x}, \bm \xi)$ only conducts the directional smoothing along $\bm \xi$.
For a matrix $\bm \Xi := (\bm \xi_1, \ldots, \bm \xi_d)$ consisting of 
$d$ orthonormal vectors, we can define $d$ directional derivatives like those in Eq.~\eqref{e4} and assemble our DGS gradient as 
%
%
\begin{equation}\label{dev_smooth_func}
{\nabla}_{\sigma, \bm \Xi}[F](\mathbf{x}) := \Big[\mathscr{D}[G_{\sigma}(0 \, |\, \mathbf{x}, \bm \xi_1)], \cdots, {\mathscr{D}}[G_{\sigma}(0\, |\, \mathbf{x}, \bm \xi_d)]\Big]\, \bm \Xi,
\end{equation}
where the orthogonal system $\bm \Xi$ and the smoothing radius $\sigma$ can be adjusted during an optimization process.




The next step is to develop an accurate DGS estimator. 
We exploit that each component of ${\nabla}_{\sigma, \bm \Xi}[F](\mathbf{x})$ only involves a 1D integral, such that the Gauss-Hermite quadrature rule \cite{2013JSV...332.4403B,Handbook} can be used to approximate the integrals with high accuracy (shown in Eq.~\eqref{GH_error}). By doing a simple change of variable in Eq.~\eqref{e4}, the GH rule can be directly used to obtain the following estimator for each directional derivative $\mathscr{D}[G_{\sigma}(0 \,|\, \mathbf{x}, \bm \xi)]$ in Eq.~\eqref{e4}
%
\begin{align}
 \widetilde{\mathscr{D}}^M[G_\sigma(0 \, | \, \mathbf{x}, \bm \xi)]  
   =
      \frac{1}{\sqrt{\pi}\sigma} \sum_{m = 1}^M w_m \,F(\mathbf{x} + \sqrt{2}\sigma v_m \bm \xi)\sqrt{2}v_m, \label{e8}
\end{align}
where $\{v_m\}_{m=1}^M$ are the roots of the $M$-th order Hermite polynomial
 and $\{w_m\}_{m=1}^M$ are quadrature weights.  
Both $v_m$ and $w_m$ can be found online\footnote{Nodes and weights for GH quadrature: \url{https://keisan.casio.com/exec/system/1281195844}} or in \cite{Handbook}. Compared with MC sampling, the error of Eq.~\eqref{e8} can be bounded by 
\begin{align}
\label{GH_error}
\hspace{-0.1cm}\big|(\widetilde{\mathscr{D}}^M- \mathscr{D})[G_\sigma] \big| \le C\frac{M\,!\sqrt{\pi}}{2^M(2M)\,!} \sigma^{2M-1}, 
\end{align}
where $M!$ is the factorial of $M$ and the constant $C>0$ is independent of $M$ and $\sigma$. 
Applying the GH quadrature rule $\widetilde{\mathscr{D}}^M$ to each component of ${\nabla}_{\sigma, \bm \Xi}[F](\mathbf{x})$ in Eq.~\eqref{dev_smooth_func}, we define the following estimator: 
\begin{equation}\label{e5}
 \widetilde{\nabla}^M_{\sigma, \bm \Xi}[F](\mathbf{x}) = \Big[\widetilde{\mathscr{D}}^M[G_{\sigma}(0 \, |\, \mathbf{x}, \bm \xi_1)], \cdots, \widetilde{\mathscr{D}}^M[G_{\sigma}(0\, |\, \mathbf{x}, \bm \xi_d)]\Big]\, \bm \Xi. 
\end{equation}
The DGS estimator has the following features: 

\begin{itemize}
    \item {\em Nonlocality}: The directional smoothing allows for a large radius $\sigma$ to capture global structures of loss landscapes and help escape from local minima. 
    \item {\em Accuracy}: The GH quadrature with the error bounded in Eq.~\eqref{GH_error} provides an estimator having much higher accuracy than MC, even when a large smoothing radius $\sigma$ is used.
    \item {\em Portability}: The DGS gradient can be integrated into a majority of gradient-based algorithms, e.g., gradient descent, Adam, and those with constraints.
    \item {\em Scalability}: The DGS estimator in Eq.~\eqref{e5} requires $M\times d$ evaluations of $F(\mathbf{x})$, and these evaluations are completely parallelizable as those in random sampling. 
\end{itemize}


\subsection{A mathematical example}
\begin{figure}[h!]
     \centering
  \includegraphics[scale = 0.45]{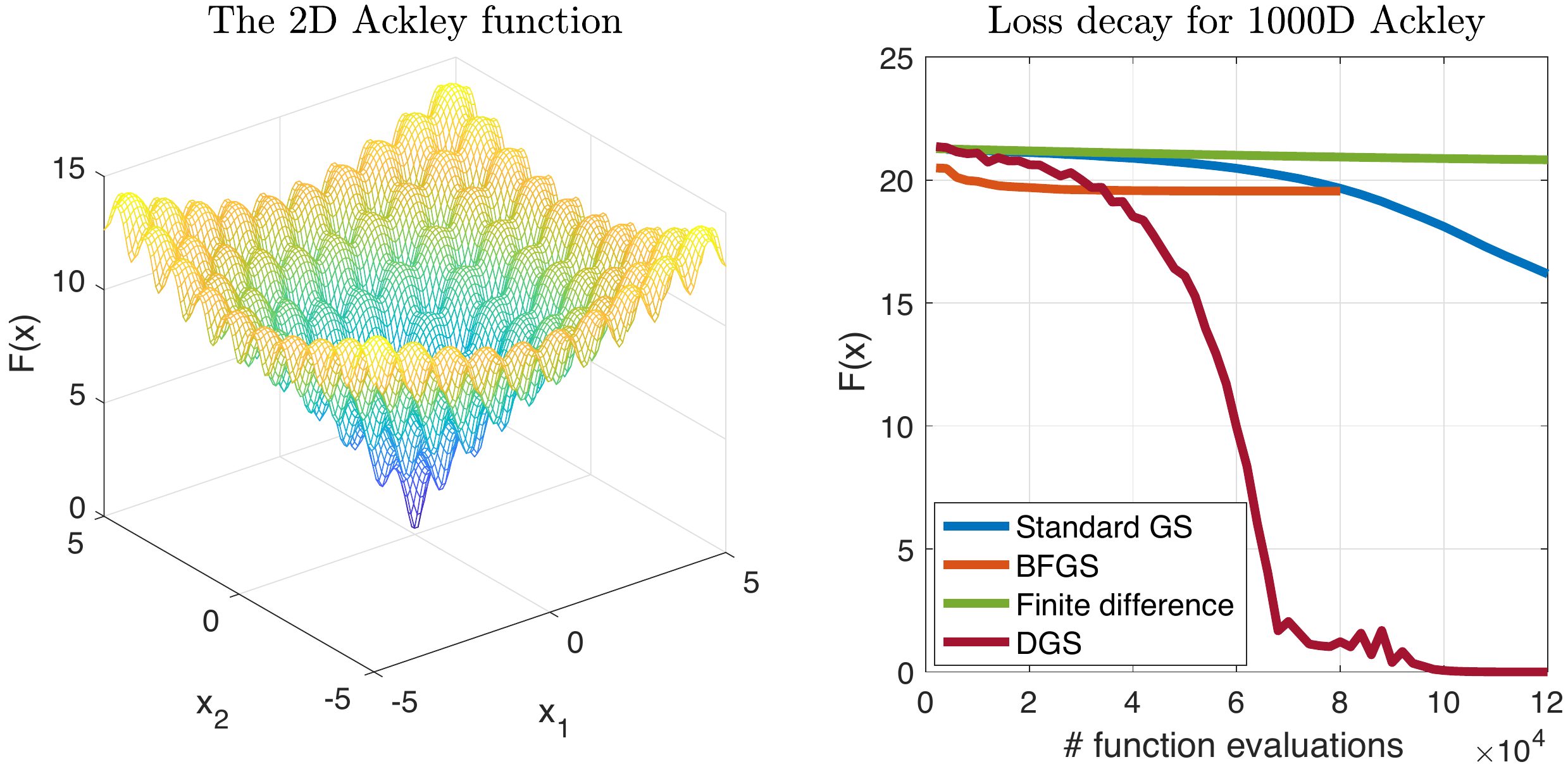}
    \caption{(Left) The landscape of the 2D Ackley function that possesses many local minima. (Right) Comparison of the loss decay w.r.t.~\# function evaluations for the 1000D Ackley function. Each curve was generated by averaging 20 independent trials with random initial states. The global minimum is $F(\mathbf{x})=0$. DGS gradient successfully found the global minimum while the other baselines are trapped in local minima.} 
    \label{fig_func}
\end{figure}

To illustrate the performance of the DGS gradient, we combine the DGS gradient with the standard gradient descent algorithm to optimize the 1000D Ackley function, which is one of the benchmark functions used to test non-convex optimization algorithms \cite{10.1145/1830761.1830794,Jamil2013ALS}. The Ackley function is defined by
\begin{equation}
F(\mathbf{x}) = -a\exp\left(-b\sqrt{\frac{1}{d}\sum_{i=1}^d x_i^2} \right)-\exp\left( \frac{1}{d}\sum_{i=1}^d \cos(c x_i)\right) + a + \exp(1),
\end{equation}
where $d$ is the dimension and $a=20, b= 0.2, c=2\pi$ are used in our experiments. The input domain $\mathbf{x} \in [-32.768, 32.768]$. The global minimum is $f(\mathbf{x}^*) = 0$, at $\mathbf{x}^* = (0,...,0)$. The Ackley function represents {non-convex} landscapes with {nearly flat outer region}.  The function poses a risk for optimization algorithms, particularly hill-climbing algorithms, to be trapped in one of its many local minima. At each iteration, we update the state $\mathbf{x}_t$ to $\mathbf{x}_{t+1}$ by
\begin{equation}
  \mathbf{x}_{t+1} = \mathbf{x}_{t} - \lambda_t \widetilde{\nabla}^M_{\sigma_t, \bm \Xi}[F](\mathbf{x}_t),     
\end{equation}
where $\bm \Xi=\mathbf{I}_d$. The learning rate $\lambda_t$ follows a polynomial decay schedule 
$\lambda_t = (\lambda_0-\lambda_T) \left(1-\frac{t}{T}\right)^\tau + \lambda_T$ with $\lambda_0 = 4000$, $\lambda_T= 0.001$, $\tau = 4$, $T = 60$. The smoothing radius also follows a polynomial decay schedule $\sigma_t = (\sigma_0-\sigma_T) \left(1-\frac{t}{T}\right)^\nu + \sigma_T$ with $\sigma_0 = 2.0$, $\sigma_T = 0.001$, $\nu = 2.0$.
We compare our method with the standard Gaussian smoothing method (i.e., replacing $\widetilde{\nabla}^M_{\sigma_t, \bm \Xi}[F](\mathbf{x}_t)$ with Eq.~\eqref{e40}), the BFGS method, and the finite difference method for estimating local gradients. The result is shown in Figure \ref{fig_func}. As shown in Figure \ref{fig_func}(Left), the Ackley function has many local minima which pose significant challenges for optimization. Figure \ref{fig_func} (Right) shows that the DGS gradient exploited its nonlocal exploration ability to skip the local minima and converge to the global minimum. The other baseline methods do not converge because they are trapped in some local minima. 

{\color{black}
\subsection{High-dimensional benchmark function demonstration}
We further test the performance of DGS method on three 2000 dimensional benchmark functions for global optimization, i.e. rotated Ellipsoidal, rotated Ackely and rotated Schaffer. Their definitions and properties can be found in \cite{finck2010real}. These rotated functions that are more challenging can be used to verify the performance and rotation-invariant property of the DGS method. To illustrate the merts of the DGS method, we provide a comparison between the DGS method and the state-of-the-art approaches on computational efficiency and accuracy through these benchmark high-dimensional functions. The compared optimization algorithms are listed as follows: (a) {\bf ES-Bpop}: the standard OpenAI evolution strategy (ES) in \cite{SHCS17} with a big population (i.e., using the same number of samples as DGS method), (b) {\bf ASEBO}: Adaptive ES-Active Subspaces for Blackbox Optimization \cite{choromanski2019complexity} with a populatioin of size $4+3\log(d)$ where $d$ is the dimensionality, (c) {\bf IPop-CMA}: the restart covariance matrix adaptation evolution strategy (CMA-ES \cite{hansen2001completely}) with increased population size \cite{auger2005restart}, (d) {\bf Nesterov}: the random search method in \cite{nesterov2017random}, (e) {\bf FD}: the classical central difference scheme, (f) {\bf Cobyla}: the Constrained Optimization BY Linear Approximation algorithm \cite{powell2007view}, (g) {\bf Powell}: the Powell's conjugate direction method \cite{powell2007view}, (h) {\bf DE}: the Differential Evolution algorithm \cite{price2006differential} and (i) {\bf PSO}: the Particle Swarm Optimization algorithm \cite{kennedy1995particle}. 

\begin{figure}[h!]
     \centering
  \includegraphics[width=1.0\textwidth]{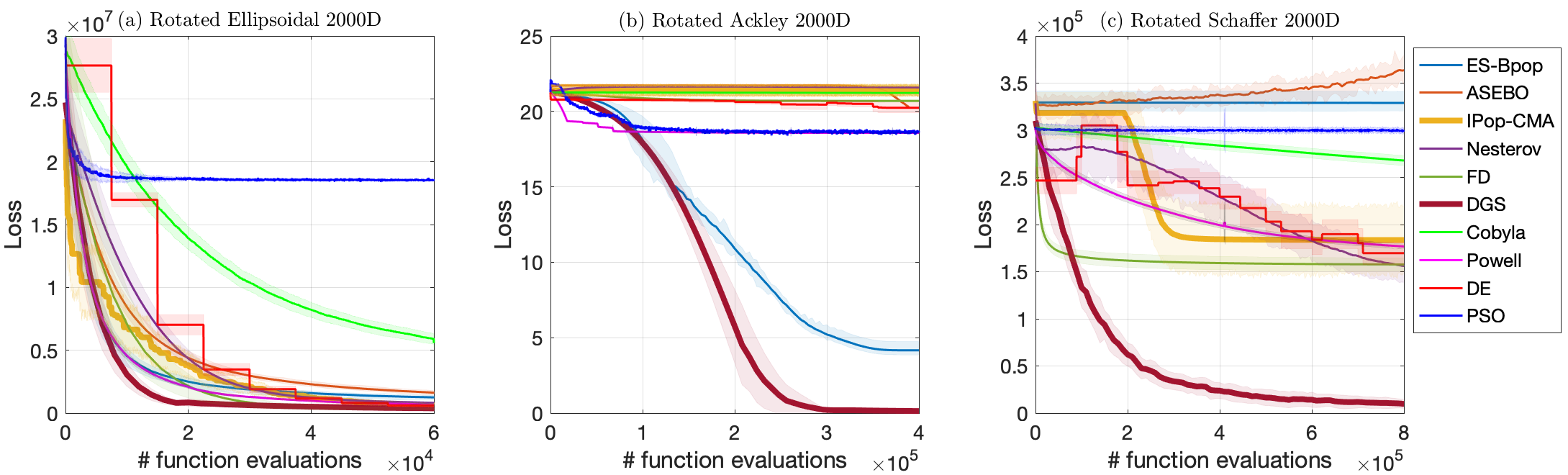}
    \caption{Performance comparison of the loss decay with respect to the number of functional evaluations for the three benchmark functions in 2000-dimensional spaces. The global minimum is 0 for all the three functions. } 
    \label{fig:rotated_fun}
\end{figure}

Figure \ref{fig:rotated_fun} shows the performance comparison for 2000D rotated functions given a same number of function evaluations. For each function, we tested 5 random rotations and we run 5 trials with different random initial states , i.e., 25 trials in total. In Figure \ref{fig:rotated_fun}, the solid lines represent the mean loss decay and the shadow areas cover the interval between the maximum and minimum loss values. The statistical confidence bounds reflect the effect of random initial states on the optimization performance. It is observed that the DGS method outperform all other baseline methods. In particular, the DGS method demonstrates significantly superior performance in optimizing the high non-convex and multi-modal functions (i.e., Ackley and Schaffer). This is because the two advantages of DGS: strong nonlocal exploration and smaller variation of gradient estimators. 
}

\section{Inverse design of wavelength demultiplexer}
In this section, we use the DGS-based nonlocal optimization method to design wavelength demultiplexer in 3D. This example is a canonical benchmarking demonstration for inverse design in nanophotonic devices. We first provide a problem description with an objective function definition and parameterization scheme. Then a methodology workflow illustrates how to incorporate DGS gradient optimization for inverse design. Finally, we demonstrate the superior performance of the optimized device using the nonlocal optimization method, investigate the effect of random initialization on robustness, and conduct constrained optimization with a limited amount of materials.

\subsection{Problem description}
\label{sec:4.1}
\begin{figure}[!ht]    
    \centering
\includegraphics[width=0.6\textwidth]{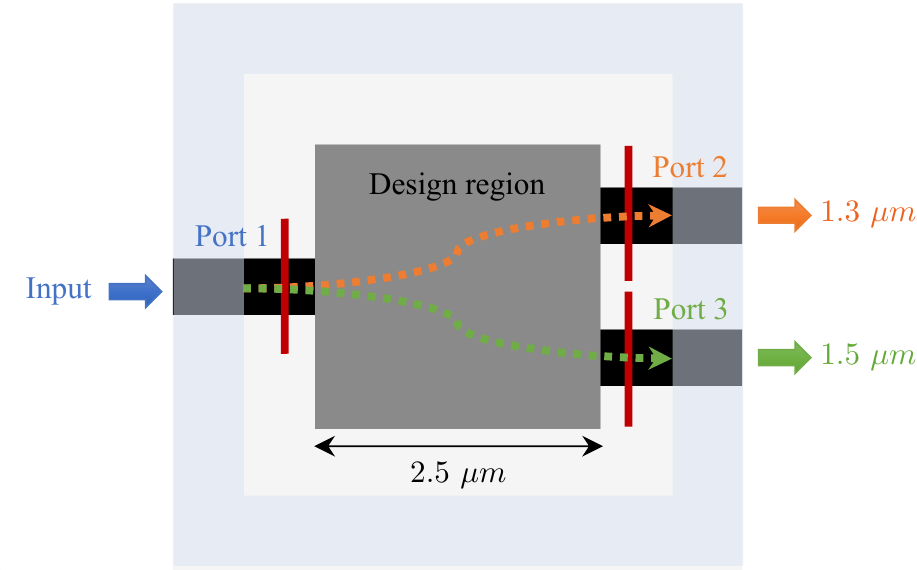}
  \caption{Illustration of wavelength demultiplexer design. The structure consists of one input waveguide (port 1), two output waveguides (port 2 and port 3), and a 2.5 $\mu m$ $\times$ 2.5 $\mu m$ design region. All three waveguides are the same, with a width of 500 nm. The outer hatched light blue frame represents the simulation domain, specifically, the perfectly matched layer (PML) boundaries. The goal of inverse design is route to 1.3 $\mu m$ through the top waveguide and 1.5 $\mu m$ through the bottom waveguide. } \label{fig:demo}
\end{figure}

As shown in Figure~\ref{fig:demo}, we choose a three-port structure with 500 nm input waveguide and output waveguides and a square 2.5 $\mu m$ $\times$ 2.5 $\mu m$ design region. We design a device for the 220 nm silicon-on-insulator (SOI) platform where the structure is constrained to a single fully etched Si layer on a SiO$_2$ substrate with air cladding. For illustration, the refractive indices of $n_{\textup{air}}=1$, $n_{\textup{SiO}_2}=1.45$ and $n_{\textup{Si}}=3.5$ are used. The purpose of inverse design is to separate 1300 nm signal to the upper waveguide and 1500 nm signal to the bottom waveguide. 

\begin{figure}[!ht]    
    \centering
\includegraphics[width=0.9\textwidth]{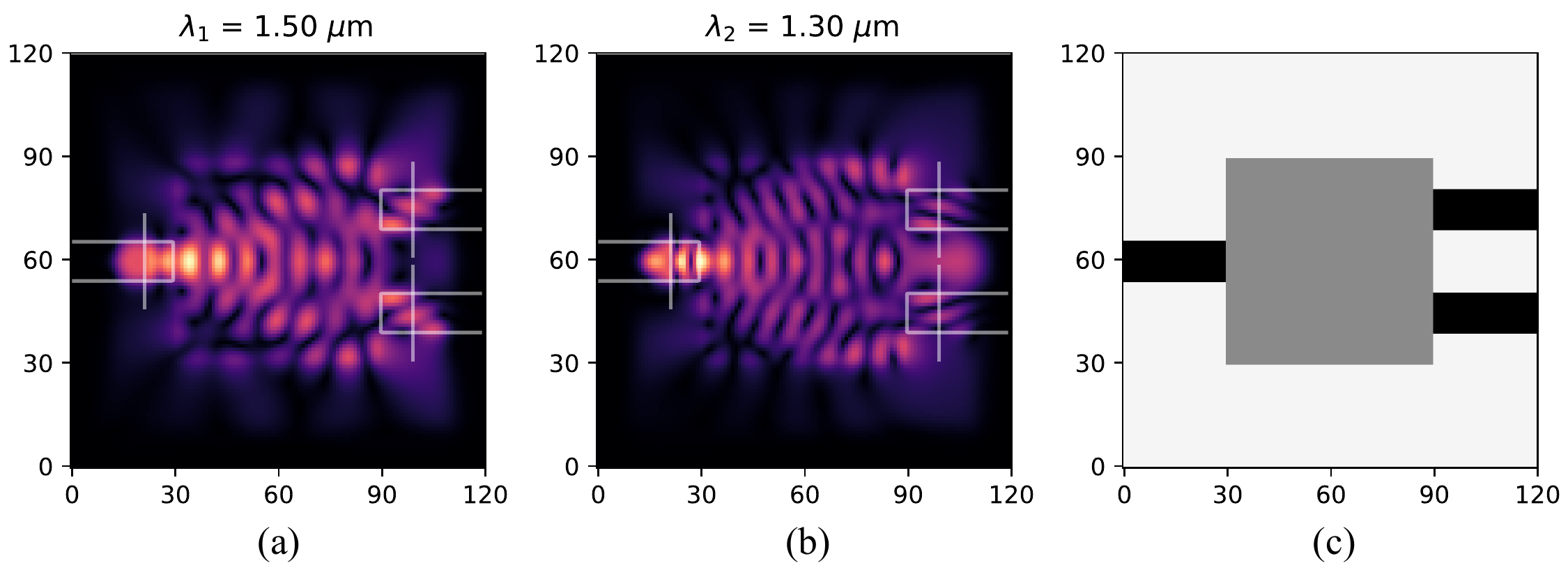}
  \caption{Initialization of inverse design. (a) Electric field intensity at 1500 nm, (b) electric field intensity at 1300 nm, and (c) initial permittvity distribution}
\label{fig:ini_design}
\end{figure}

In this example, the fundamental first-order mode of the input waveguide is used as the input mode for the inverse design, and the fundamental first-order modes of the two output waveguides are used as the output modes. Initially, the permittivity in design region is homogeneously distributed as shown in Figure~\ref{fig:ini_design} (c) and the resulting electric field intensity $E_{z_1}$ at 1500 nm and $E_{z_2}$ at 1300 nm are calculated by FDFD simulations, as shown in Figure~\ref{fig:ini_design} (a) and (b) respectively. To conduct the FDFD simulation, {\color{black} the computational domain of entire structure $\mathbf{x}$, as shown in Eq.~\eqref{eq:obj} is discretized by 120 $\times$ 120 pixels and the design region is parametrized by 60 $\times$ 60 pixels, leading to the pixelated design.} For ease of inverse design process, we define a relative permittivity $\bm{\epsilon}_i$ with a minimum value $\bm{\epsilon}_{\min} = 1.0$ (white color in Figure~\ref{fig:ini_design} (c) ) and maximum value $\bm{\epsilon}_{\max}=12.0$ (black color in Figure~\ref{fig:ini_design} (c)). The initial permittivity distribution with $\bm{\epsilon}_{\textup{ini}}=6.5$ are set up for the design region.  


\subsection{Objective function}
We define the output modes of interest as $\zeta_1$ and $\zeta_2$ over output surface $S$. The device performance is then specified by the overlap integral, which is given by 
\begin{equation}
    \mathbf{c}^{\dagger }\mathbf{E}_i (\bm{\epsilon}(\mathbf{x})) = \iint_{S} \zeta_i \cdot \mathbf{E}_i(\bm{\epsilon}(\mathbf{x})) d S \label{eq:integral}
\end{equation}
where we use it to compute the mode coupling efficiency into each output mode. To achieve the goal of maximum overlap integral, the optimization problem is formulated as follows: 
\begin{equation}
\begin{aligned}
    \min_{\mathbf{x}}& \quad -\exp \left[ \log(\mathbf{c}^{\dagger }\mathbf{E}_1) - \log(\mathbf{c}^{\dagger }\mathbf{E}_{z_1})+ \log(\mathbf{c}^{\dagger }\mathbf{E}_2)  - \log(\mathbf{c}^{\dagger }\mathbf{E}_{z_2})  \right]
     \\
    \textup{subject~to} & \quad   \nabla \times \frac{1}{\mu}\nabla  \times \mathbf{E}_1 - \omega_1^2 \bm{\epsilon} (\mathbf{x}) \mathbf{E}_1 = -i \omega_1 \mathbf{J}_1  \\
    & \quad \nabla \times \frac{1}{\mu}\nabla  \times \mathbf{E}_2 - \omega_2^2 \bm{\epsilon} (\mathbf{x}) \mathbf{E}_2 = -i \omega_2 \mathbf{J}_2 \label{eq:obj}
\end{aligned}
\end{equation}
where $\omega_1$ and $\omega_2$ are the angular frequencies at 1300 and 1500 nm, $\mathbf{E}_1$ and $\mathbf{E}_2$ are the electric field, and $\mathbf{J}_1 $ and $\mathbf{J}_1 $ inject input sources into the waveguide for frequency $\omega_1$ and $\omega_2$. The objective is a sum with four terms using a negative log-sum-exp smooth approximation of the maximum function, and each term corresponds to a sub-objective. As shown in Eq.~\eqref{eq:obj}, two terms $\log(\mathbf{c}^{\dagger }\mathbf{E}_1)$ and $\log(\mathbf{c}^{\dagger }\mathbf{E}_2)$ correspond to maximizing transmission efficiency through the top waveguide at 1300 nm and bottom waveguide at 1500 nm, given the specific permittivity distribution $\bm{\epsilon}(\mathbf{x})$, and {\color{black} two terms $ \log(\mathbf{c}^{\dagger }\mathbf{E}_{z_1})$ and $ \log(\mathbf{c}^{\dagger }\mathbf{E}_{z_2})$ correspond to the initial overlap integral given 
the homogeneous permittivity distribution $\bm{\epsilon}_{\textup{ini}}(\mathbf{x}_{\textup{ini}})$. Note that, the $\mathbf{E}_{z_1}$ and $\mathbf{E}_{z_2}$ are constant during the design process,  which are mainly used to normalize the objective, but the $\mathbf{E}_1$ and $\mathbf{E}_2$ vary along with the update of $\mathbf{x}$ to minimize the objective function.} One may consider to include other sub-objectives using alternative mathematical formulation for improving the optimization performance. Interested reader may find more discussion in \cite{Su2020}. 

\subsection{Parameterization scheme}
\begin{figure}[!ht]    
    \centering
\includegraphics[width=0.98\textwidth]{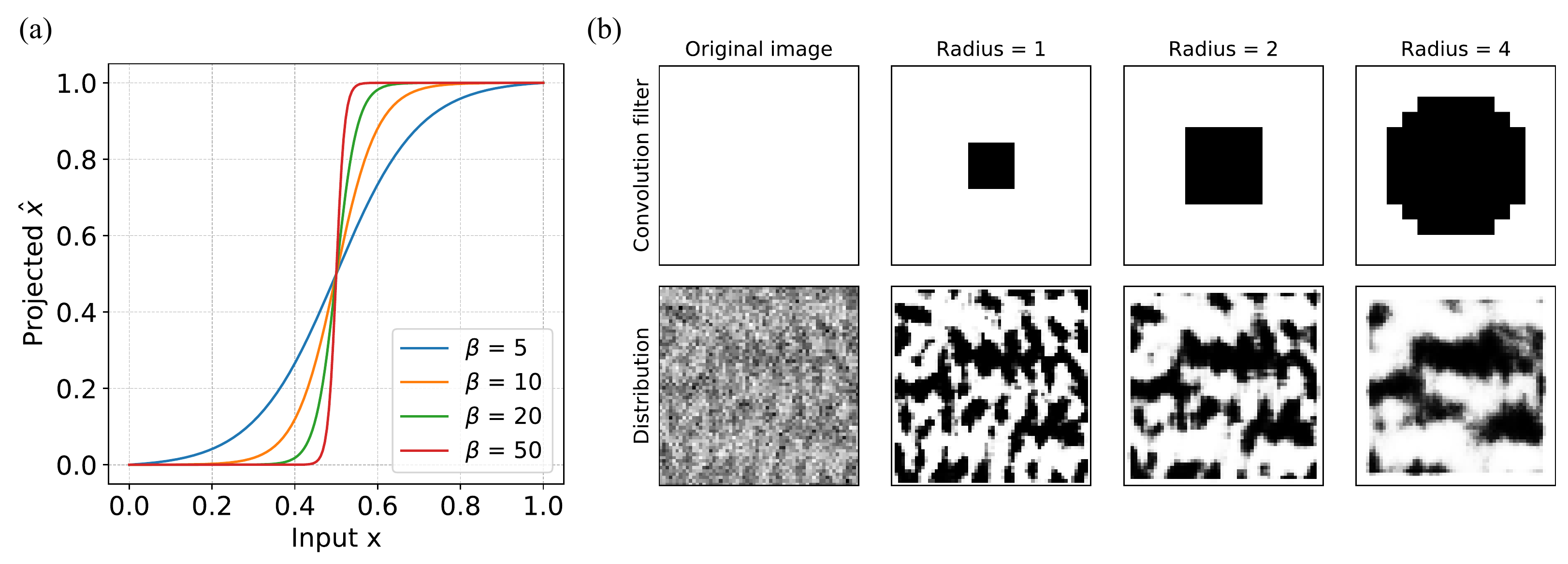}
  \caption{Parameterization scheme for inverse design. (a) Nonlinear projection function for binarizing the input design variables and (b) visualization of convolution filtering with different filter radius given a specific projection strength.}
\label{fig:para}
\end{figure}

By selecting an appropriate parameterization, the fabrication constraints in optimization can be naturally imposed. The parameterization scheme here consists of two crucial operators: nonlinear projection and convolution filtering. Nonlinear projection aims to binarize the permittivity distribution by 
\begin{equation}
      \bm{\epsilon} (\mathbf{x}) =  \bm{\epsilon}_{\min} + (\bm{\epsilon}_{\max} - \bm{\epsilon}_{\min}) \varphi(\mathbf{x}), \quad \varphi(\mathbf{x}) = \frac{\tanh (\beta \cdot \eta) + \tanh(\beta \cdot (\mathbf{x} - \eta))}{\tanh(\beta \cdot \eta) + \tanh(\beta \cdot (1-\eta))} \label{eq:projection}
\end{equation}
where $\beta$ is the coefficient of projection strength, and $\eta$ is the center of the projection. Figure~\ref{fig:para} (a) shows the nonlinear projected mapping between original input $\mathbf{x}$ and projected $ \hat{\mathbf{x}} = \varphi(\mathbf{x})$ given different projection strength and fixed $\eta=0.5$. As the increasing of projection strength $\beta$, the projected $\hat{\mathbf{x}}$ shows a clear trend to binary value 0 or 1.  

The convolution operator is used as a blurring filter that results in smooth features of the permittivity distribution and avoids the tiny features that are less than the minimum feature size of fabrication. Integrating nonlinear projection ($\beta=50$ and $\eta=0.5$) and convolution filtering, we visualize the parameterized distribution with varying circle radius based on a specific permittivity distribution, as shown in Figure~\ref{fig:para} (b). It is clear to see that the parameterized distribution shows a clear black-white pixel (material layout) without intermediate grey pixels. The feature size can be controlled by determining a specific convolution radius.

\subsection{Methodology workflow}

We illustrate a workflow to implement DGS-based nonlocal optimization method for inverse design problem. Figure~\ref{fig:workflow} shows the four core components, that are parameterization, physics simulation, objective formulation and optimization. The detailed procedure is summarized as follows: 

\begin{figure}[!ht]    
    \centering
\includegraphics[width=0.98\textwidth]{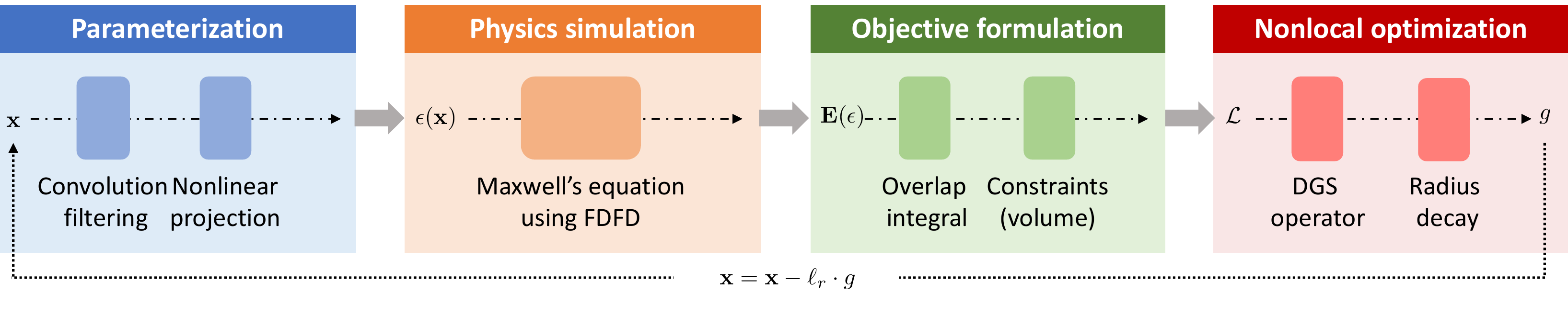}
  \caption{Methodology workflow of inverse design using nonlocal optimization method.}
\label{fig:workflow}
\end{figure}

\begin{itemize}
    \item Step 0: Initialization. An initial design variable $\mathbf{x}_0$ are set up through a homogeneously distribution or a random distribution with noise. 
    \item Step 1: Parameterization. For design variable $\mathbf{x}_k$ at the $k$-th iteration, convolution filtering with a specific radius is imposed to eliminate the small features, followed by the nonlinear projection in Eq.~\eqref{eq:projection} that binarizes the design variables. Parameterization builds up a transformation between design variable $\mathbf{x}_k$ and the corresponding permittivity distribution $\bm{\epsilon}({\mathbf{x}_k})$.
    
    \item Step 2: Physics simulation. The permittivity distribution $\bm{\epsilon}({\mathbf{x}_k})$ are taken as input to physics simulation, for example, electromagnetic simulation. The electric field intensity $\mathbf{E}(\bm{\epsilon}({\mathbf{x}_k}))$ are obtained by solving the Maxwell's equation in Eq.~\eqref{eq:maxwell} using FDFD method. 
    
    \item Step 3: Objective formulation. The objective function in Eq.~\eqref{eq:obj} is formulated to conduct the optimization for inverse design. The resulting $\mathbf{E}(\bm{\epsilon}({\mathbf{x}_k}))$ is used to calculate the overlap integral and then yield a scale value $\mathcal{L}_k$ as the loss. The constraints on fabrication and volume fraction are also defined in this step. 
    
    \item Step 4: Nonlocal optimization. The Step 1-3 can be considered as a forward evaluation where a set of input-output paired samples $\mathbb{S} = \left\{ (\mathbf{x}_k^{(1)}, \mathcal{L}_k^{(1)}),..., (\mathbf{x}_k^{(l)}, \mathcal{L}_k^{(l)}) \right\}$ can be drawn for DGS gradient operator. These samples drawn by Gauss-Hermite quadrature rule are used to estimate each directional derivative in Eq.~\eqref{e8} and are then assembled to accurately approximate the full $d$-dimensional gradient $\mathbf{g}_k$ in Eq.~\eqref{e5}. The DGS gradient is well-suited to update the design variable via gradient descent algorithm:
    \begin{equation}
        \mathbf{x}_{k+1} = \mathbf{x}_k - \ell_k \cdot \mathbf{g}_k \label{eq:update}
    \end{equation}
     The new design variable $\mathbf{x}_{k+1}$ goes back to Step 1 for iterative updating until the convergence criteria is satisfied. 
\end{itemize}


To improve the optimization performance on accuracy, convergence and robustness, we implement an adaptive decay scheme for updating hyperparameters, including a large DGS radius $\sigma_{r}$ in Eq.~\eqref{e8}, learning rate $\ell_r$ in Eq.~\eqref{eq:update} and projection strength $\beta$ in Eq.~\eqref{fig:para}. Specifically, the quadratic decay is used 
\begin{equation}
    z_l = (z^\textup{ini} - z^\textup{end}) \times (1.0 - \frac{k}{k_{\max}})^{\alpha} + z^{\textup{end}},
\end{equation}
where $z_l$ represents the hyperparameters that can be $\sigma_{r}$, $\ell_r$ or $\beta$ at the $k$-th iteration, $k_{\max}$ is the maximum of iteration and $\alpha=2$ is the coefficient of decay rate. $z^\textup{ini}$ and $z^\textup{end}$ are the initial value and end value of hyperparameters respectively. In this study, we use $\sigma_r^{\textup{ini}} = 0.25$ and $\sigma_r^{\textup{end}} = 0.05$ for DGS radius, $\ell_r^{\textup{ini}} = 1.0$ and $\ell_r^{\textup{end}} = 0.01$ for learning rate, and $\beta^{\textup{ini}} = 0.2$ and $\beta^{\textup{end}} = 0.05$ for the reciprocal of projection strength. In addition, the radius $r=2$ in convolution filtering is used for parameterization. The physics simulation is achieved by $\textit{ceviche}$ (\url{https://github.com/fancompute/ceviche}) that is an electromagnetic simulation tool for solving Maxwell's equations. In DGS gradient operator, five GH quadrature points are used and the nodes of points in practical computation can be reduced to three due to symmetric property \cite{2013JSV...332.4403B}. All numerical experiments (physics simulation and optimization) are implemented in Python 3.6 and conducted on a cluster with 44 Intel Xeon E5-2699 v4 CPUs at 2.20 GHz. Each iteration in optimization takes around 1.2 min using a parallel implementation of DGS gradient operator.

\subsection{Result and discussion}

\begin{figure}[!ht]    
    \centering
\subfigure[]{\includegraphics[width=0.98\textwidth]{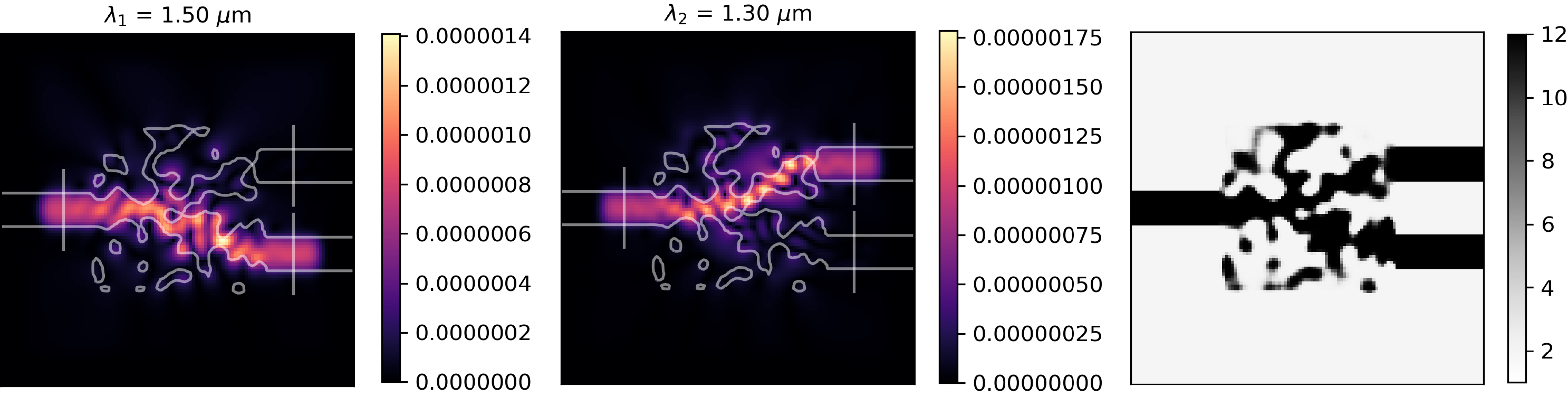}}
\subfigure[]{\includegraphics[width=0.98\textwidth]{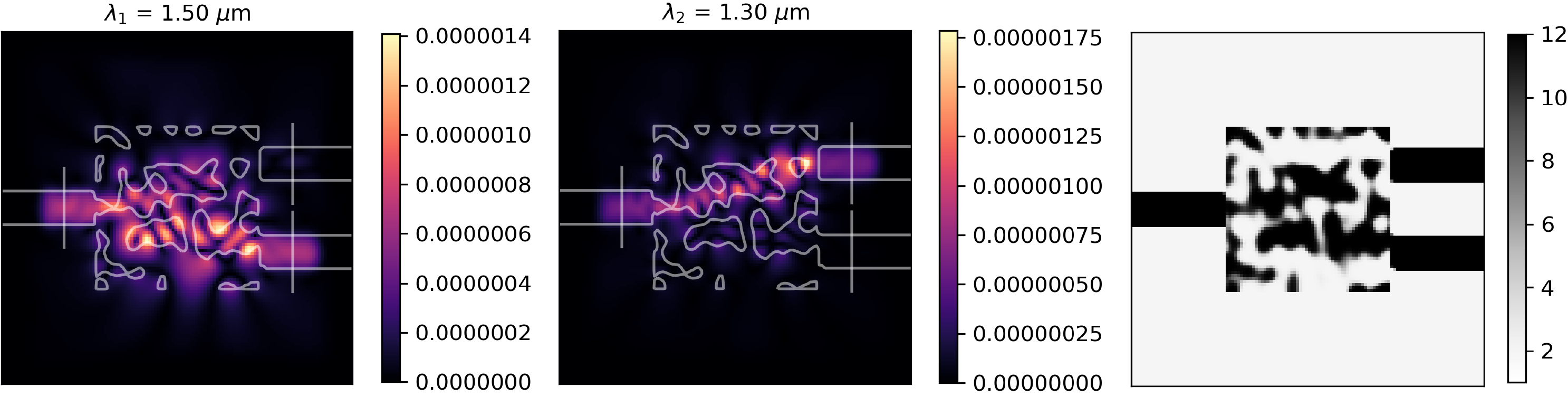}}
  \caption{Electric field intensity of the optimized device at 1500 nm (left column) and 1300 nm (middle column), as well as the optimized permittivity distribution (right column) using (a) nonlocal optimization method with DGS gradient and (b) local gradient based optimization}
\label{fig:dgs_adam}
\end{figure}

\begin{figure}[!ht]    
    \centering
\includegraphics[width=0.5\textwidth]{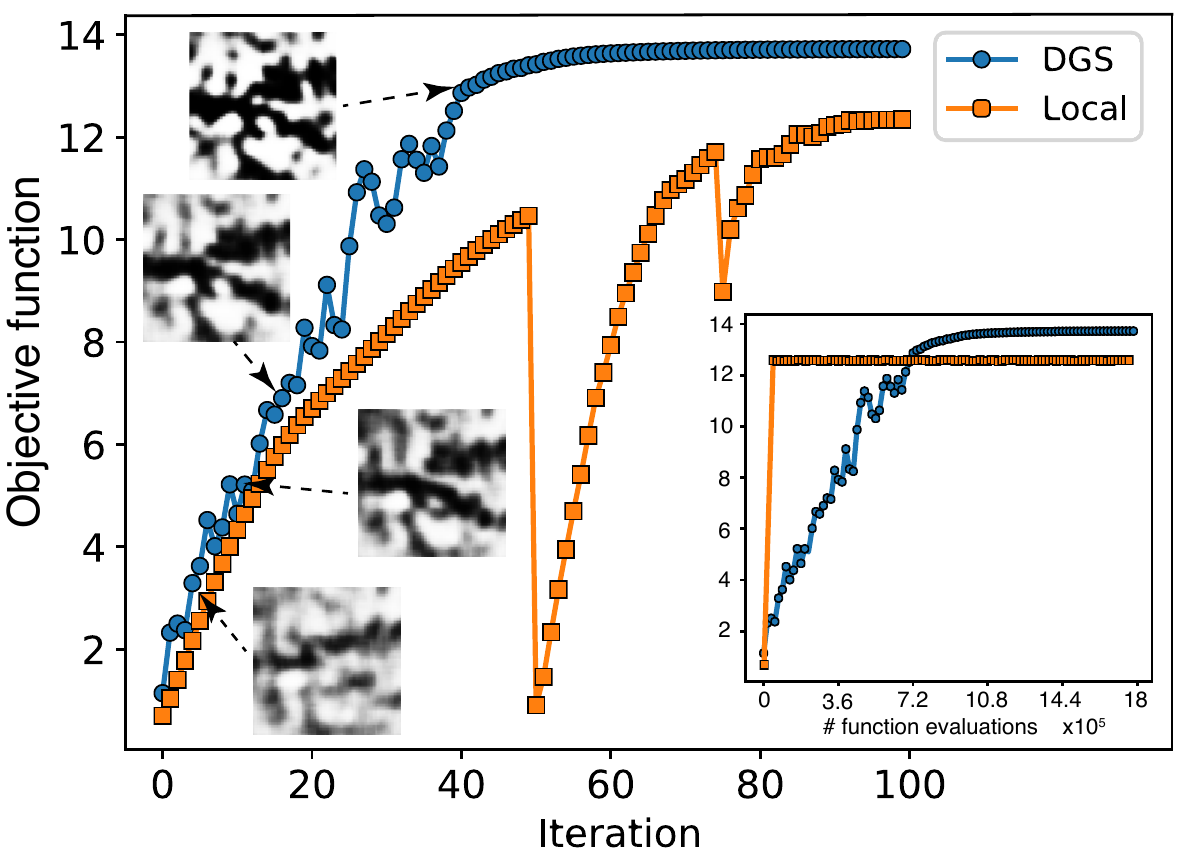}
  \caption{The optimization iteration history. The blue circles represent the optimization using DGS gradient operator and the orange squares represent the optimization using local gradient algorithm.}
\label{fig:iter}
\end{figure}

The methodology workflow of inverse design is applied here to maximize the performance of nanophotonic devices described in Section \ref{sec:4.1}. The final optimized design and the corresponding electric field intensity are diagrammed in Figure~\ref{fig:dgs_adam}. Note that the device designed by nonlocal optimization method using the DGS gradient, shown in Figure~\ref{fig:dgs_adam} (a), displays a nonintuitive geometry while retaining relatively large features and a clear permittivity distribution with ideal binarization. The light takes a relatively confined path through the structure at both wavelengths. The optimization history, shown in Figure~\ref{fig:iter}, provides iterative changes of the permittivity distribution during the optimization process. {\color{black} At each iteration, the DGS method updates the smoothing gradient estimator that requires $5\times 60\times 60$ evaluations of objective functions, which requires significantly larger computational cost than the local gradient algorithm. We realize this is the computational challenge of the DGS method and we provide a detailed discussion of computational cost and several strategies to mitigate this burden, particularly given a limited computational budget. In fact, the iteration results shown here aim to illustrate the detailed optimization process using $\textit{gradient information}$. This is different from the derivative-free global optimization algorithms, such as GA, PSO, and Bayesian optimization.} It can be seen that the local gradient method shows significant oscillations at 50 and 75 iterations, shown in Figure~\ref{fig:iter}. This is mainly due to the fact that the projection strength is increased by a discrete step function in the classical local gradient method. The nonlocal optimization method using the DGS gradient shows a relatively smoothing iteration curve since a dynamic decay mechanism is implemented to adaptively update the projection strength. {\color{black} From the computational cost perspective, we also show a fair comparison using the same number of evaluations of the objective function (see the subfigure in Figure~\ref{fig:iter} (right corner)). It is clear to note that the local gradient shows a fast convergence but it quickly traps into a local minimum and difficult to escape even though a large number of evaluations are performed. On the contrary, the DGS increases slowly initially but gradually converges to a better solution, $f_{\textup{DGS}}^* = 13.58$, that performs $\sim10\%$ improvement compared with the local gradient method that is $f_{\textup{local}}^* = 12.40$.} The gap at the final objective values between the nonlocal optimization method and the local gradient method is also demonstrated by the optimized design and the resulting electric field intensity in Figure~\ref{fig:dgs_adam}. It is easy to observe that the light path in local gradient design, specifically at 1500 nm wavelength (left in Figure~\ref{fig:dgs_adam}(b)) is relatively diffused. 


\subsubsection{Effect of random initialization with Gaussian noise}

\begin{figure}[!ht]    
    \centering
\subfigure[]{\includegraphics[width=0.3\textwidth]{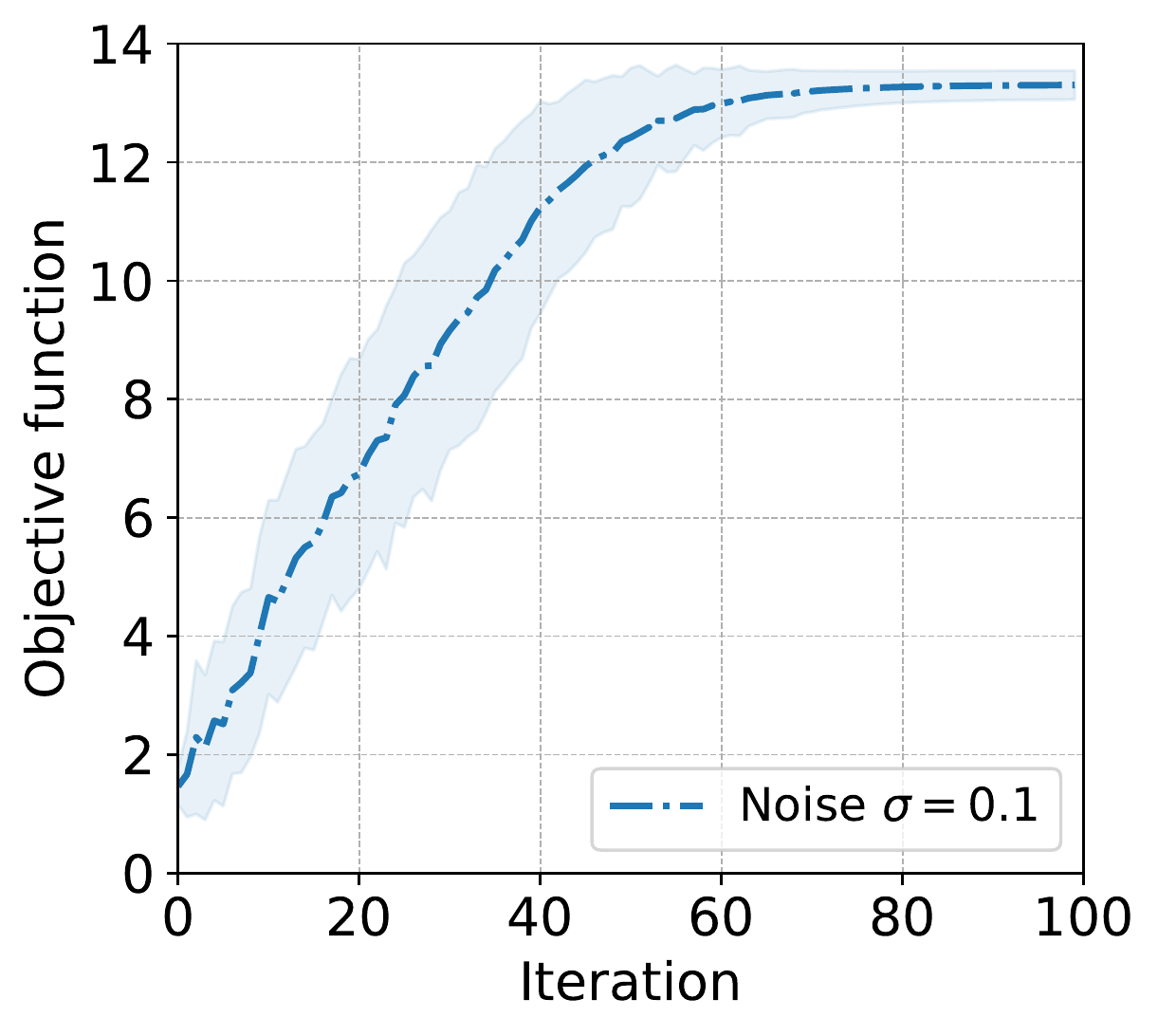}}
\subfigure[]{\includegraphics[width=0.3\textwidth]{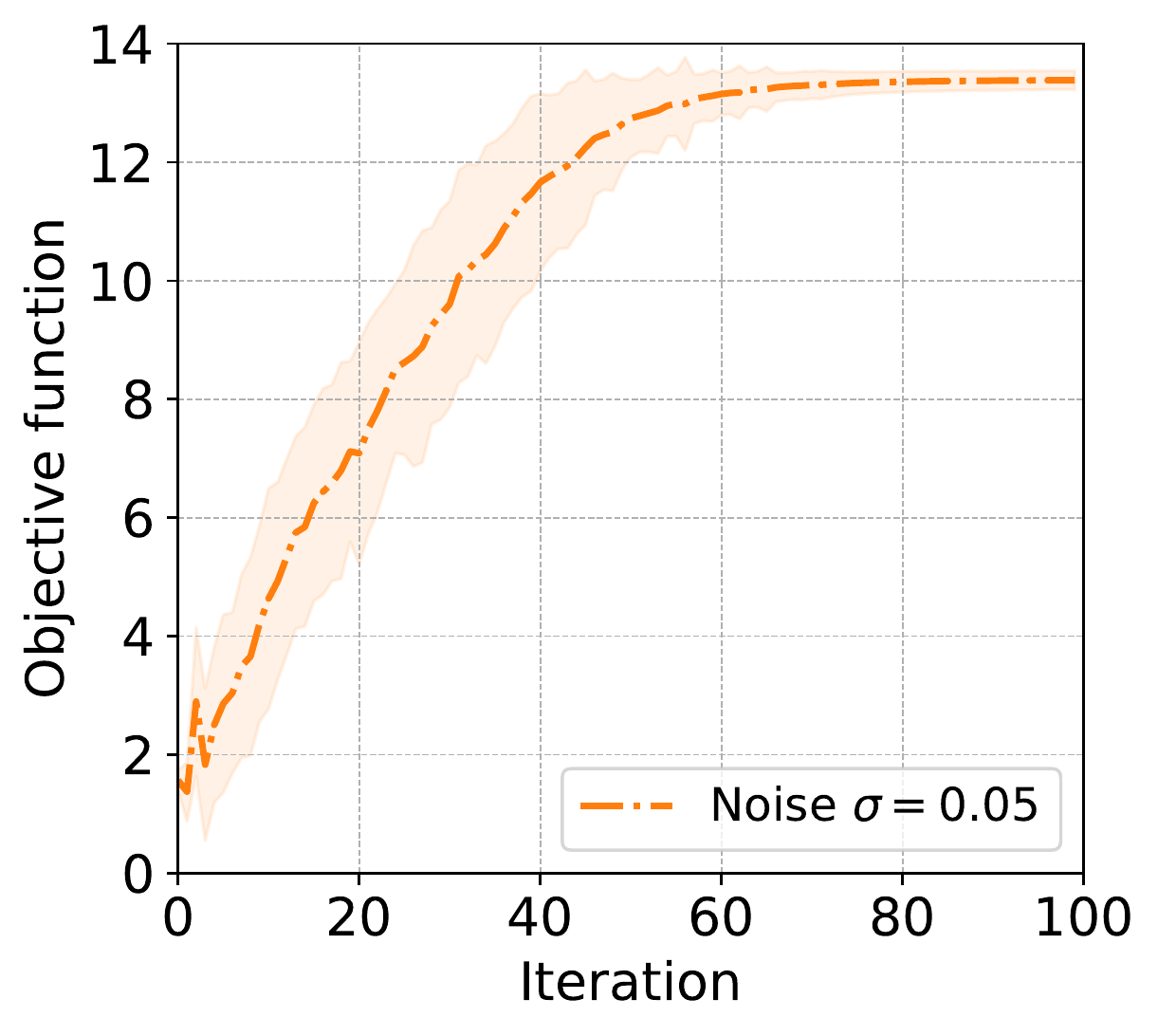}}
\subfigure[]{\includegraphics[width=0.3\textwidth]{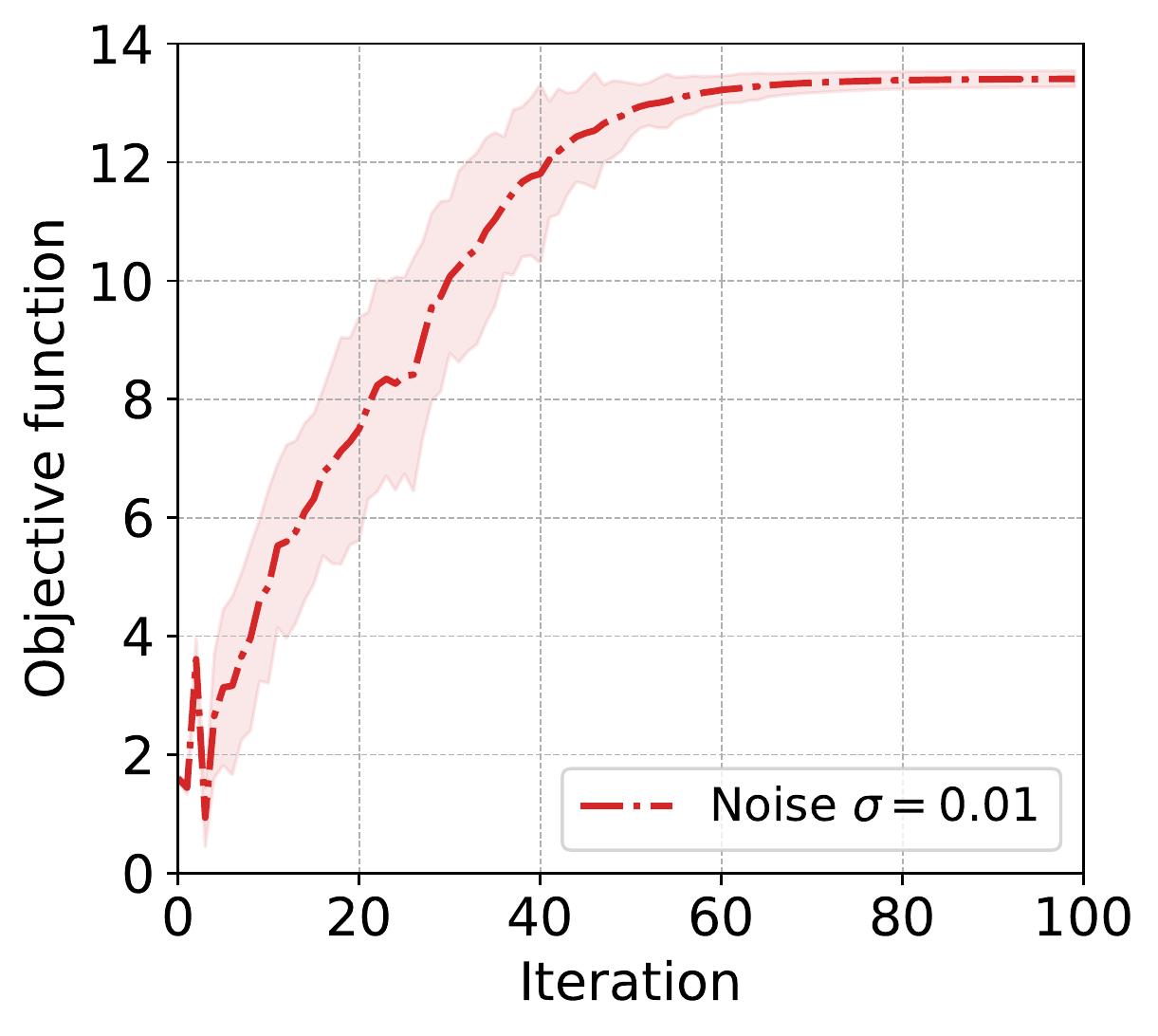}}
  \caption{Study of the local minima with three initialization with random Gaussian noise for the inverse design performance. The optimization is run 100 times with random {\color{black}initial guess} (different random seeds). (a) Gaussian noise $N(0,0.1)$, (b) Gaussian noise $N(0,0.05)$ and (c) Gaussian noise $N(0,0.01)$. }
\label{fig:CI_noise}
\end{figure}

\begin{figure}[!ht]    
    \centering
\subfigure[]{\includegraphics[width=0.3\textwidth]{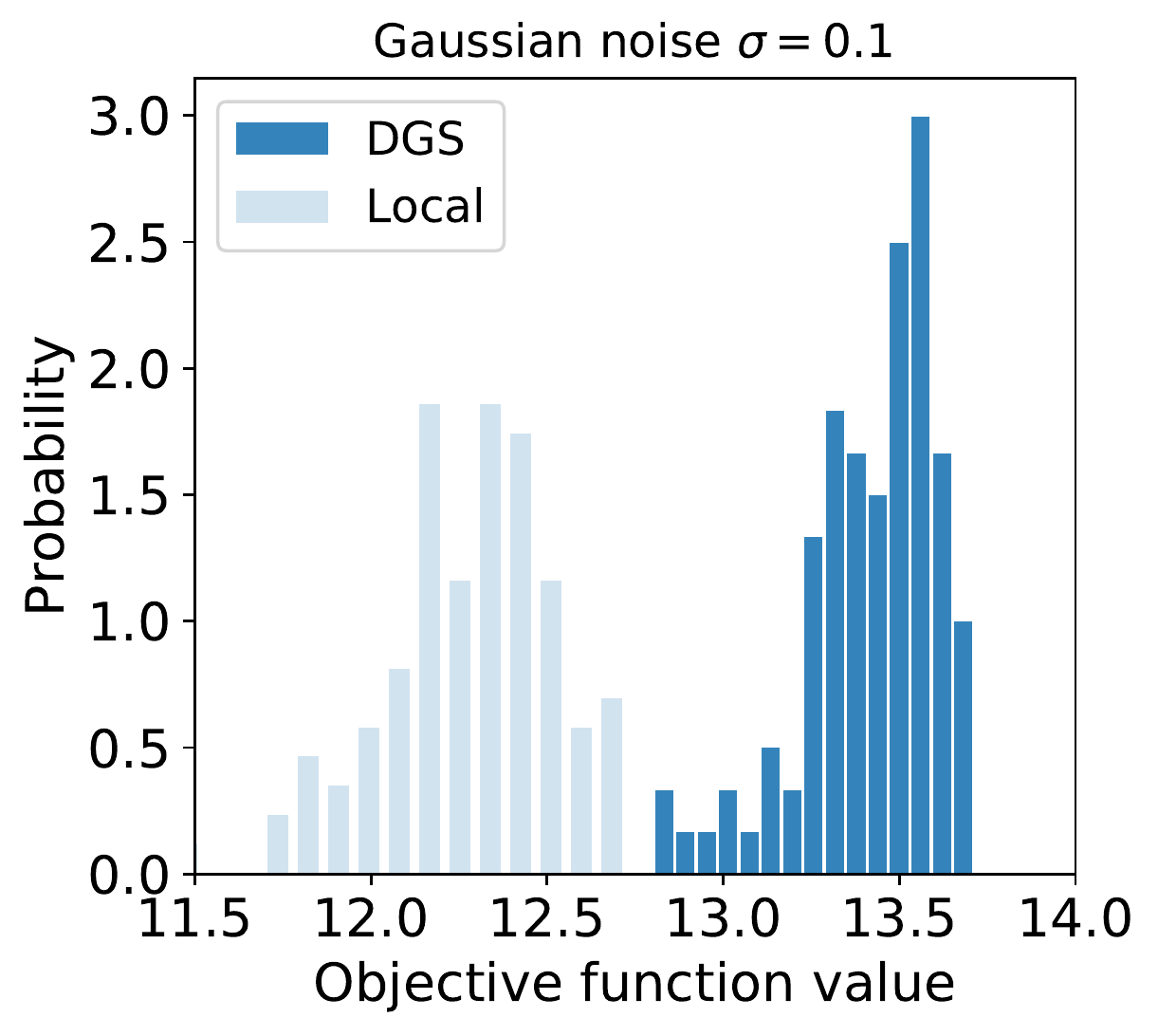}}
\subfigure[]{\includegraphics[width=0.3\textwidth]{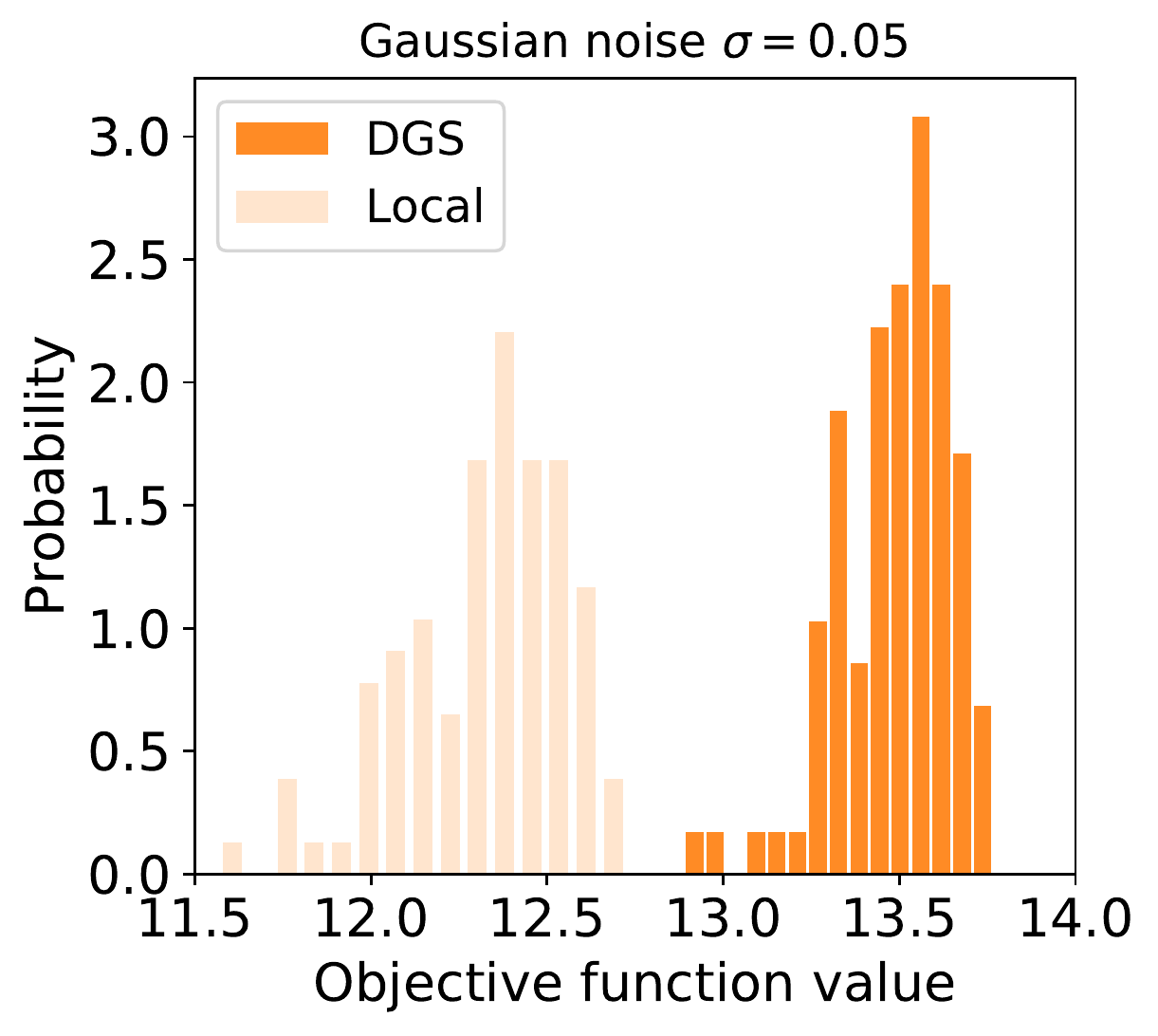}}
\subfigure[]{\includegraphics[width=0.3\textwidth]{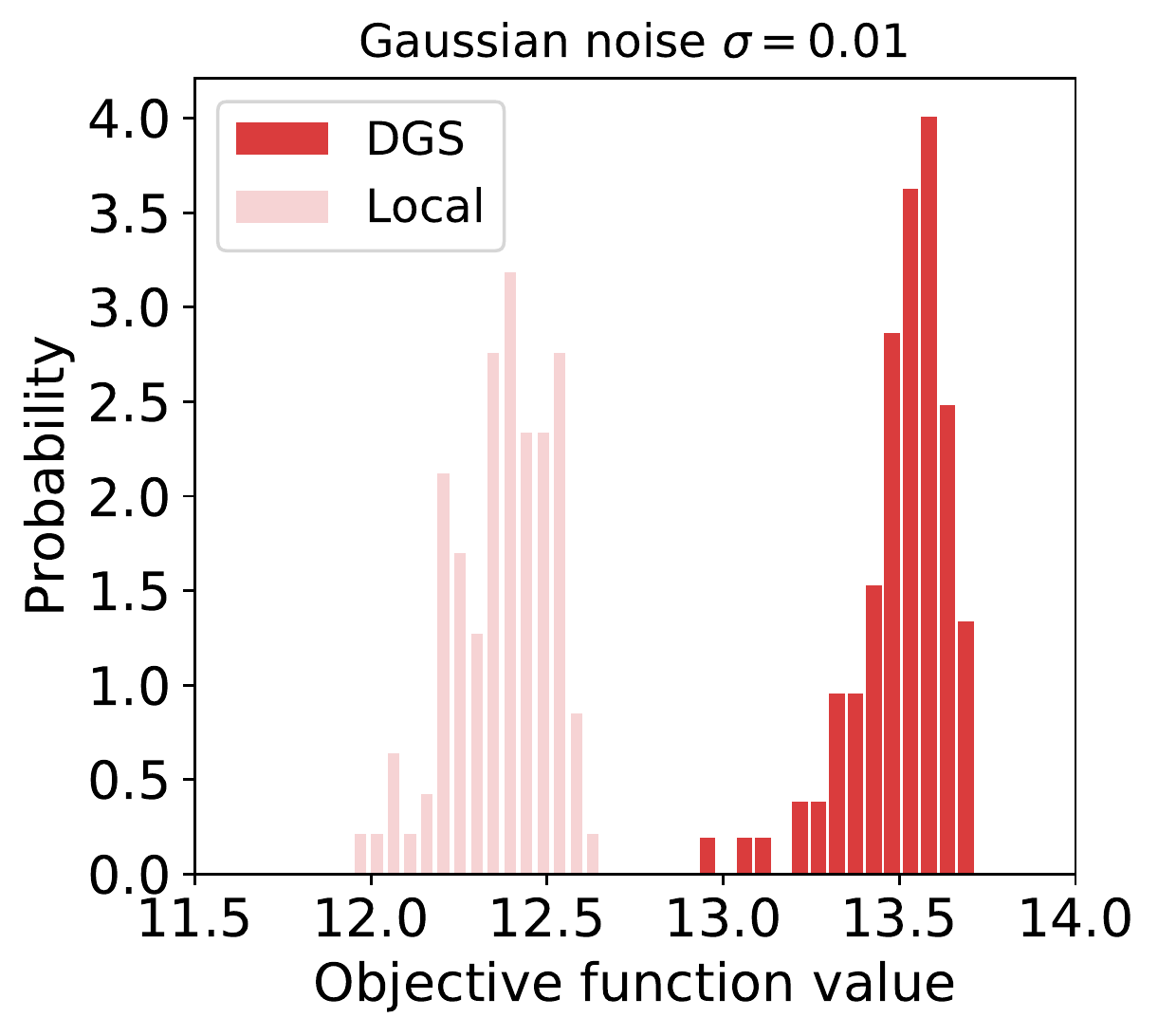}}
  \caption{ Histogram of the objective function values of 100 local minmia given different noise levels. Dark color represents the distribution using the nonlocal optimization method with DGS gradient and light color represents the distribution using local gradient algorithm. (a) Gaussian noise $N(0,0.1)$, (b) Gaussian noise $N(0,0.05)$ and (c) Gaussian noise $N(0,0.01)$. }
\label{fig:hist_noise}
\end{figure}

Here we investigate the effect of random initialization on the optimized permittivity distribution and device performance. In this case, we repeatably run 100 times optimization with three different levels of randomness described by Gaussian noise: $N(0,0.1), N(0,0.05)$ and $N(0, 0.01)$. The iteration history and histogram of final objective values are shown in Figure~\ref{fig:CI_noise} and Figure~\ref{fig:hist_noise}. The solid curves in Figure~\ref{fig:CI_noise} represent the mean value of 100 trials and the dash area represents the confidence intervals with $[-\sigma, +\sigma]$, where $\sigma$ is the standard deviation of the objective values at a specific iteration. It can be seen that the variation of objective value is relatively large but it is quickly narrowed and converged to the final value, which is pretty close in all three levels of randomness. The histogram in Figure~\ref{fig:hist_noise} shows the distribution of the final objective values using the nonlocal optimization method with DGS gradient (dark color) and local gradient (light color). The nonlocal method outperforms the local gradient in terms of the objective values and shows a smaller variation in all three levels of randomness. Table \ref{tab:t2} provides a statistical comparison of objective values between the nonlocal method using DGS gradient and local gradient methods. Compared with the local gradient method, the nonlocal optimization method shows superior performance with an improvement of 9.37\%, 9.42\%, and 9.22\% in the mean value respectively. For the standard deviation, two methods show almost the same variation level if the noise is tiny, specifically at $\sigma_N=0.01$ but the nonlocal optimization method using DGS gradient achieves a significant reduction of 21.1\% and 30.6\% when the noise level is relatively large, typically at $\sigma_N=0.1$ and $\sigma_N=0.05$. Through a statistical analysis of the objective values, the nonlocal optimization method shows a higher objective value and much smaller uncertainty on the final performance, particularly given relatively large randomness associated with the {\color{black}initial guess}. In other words, the nonlocal optimization method demonstrates stronger robustness and reliability to resist the local minima caused by random initialization. 

\begin{table}[!ht]
\centering
\caption{Statistical comparison of optimized performance between the nonlocal optimization method with DGS gradient and local gradient method}
\label{tab:t2}
\begin{tabular}{ccccc}
\hline
      & \multicolumn{2}{c}{Nonlocal optimization} & \multicolumn{2}{c}{Local gradient} \\ \hline
Gaussian noise & Mean       & Standard deviation        & Mean             & Standard deviation             \\
$\sigma_N=0.1$   & 13.42      & 0.191      & 12.29            & 0.242           \\
$\sigma_N=0.05$  & 13.48      & 0.159      & 12.33            & 0.229           \\
$\sigma_N=0.01$  & 13.51      & 0.136      & 12.37           & 0.141           \\ \hline
\end{tabular}
\end{table}

\begin{figure}[!ht]    
    \centering
\includegraphics[width=1\textwidth]{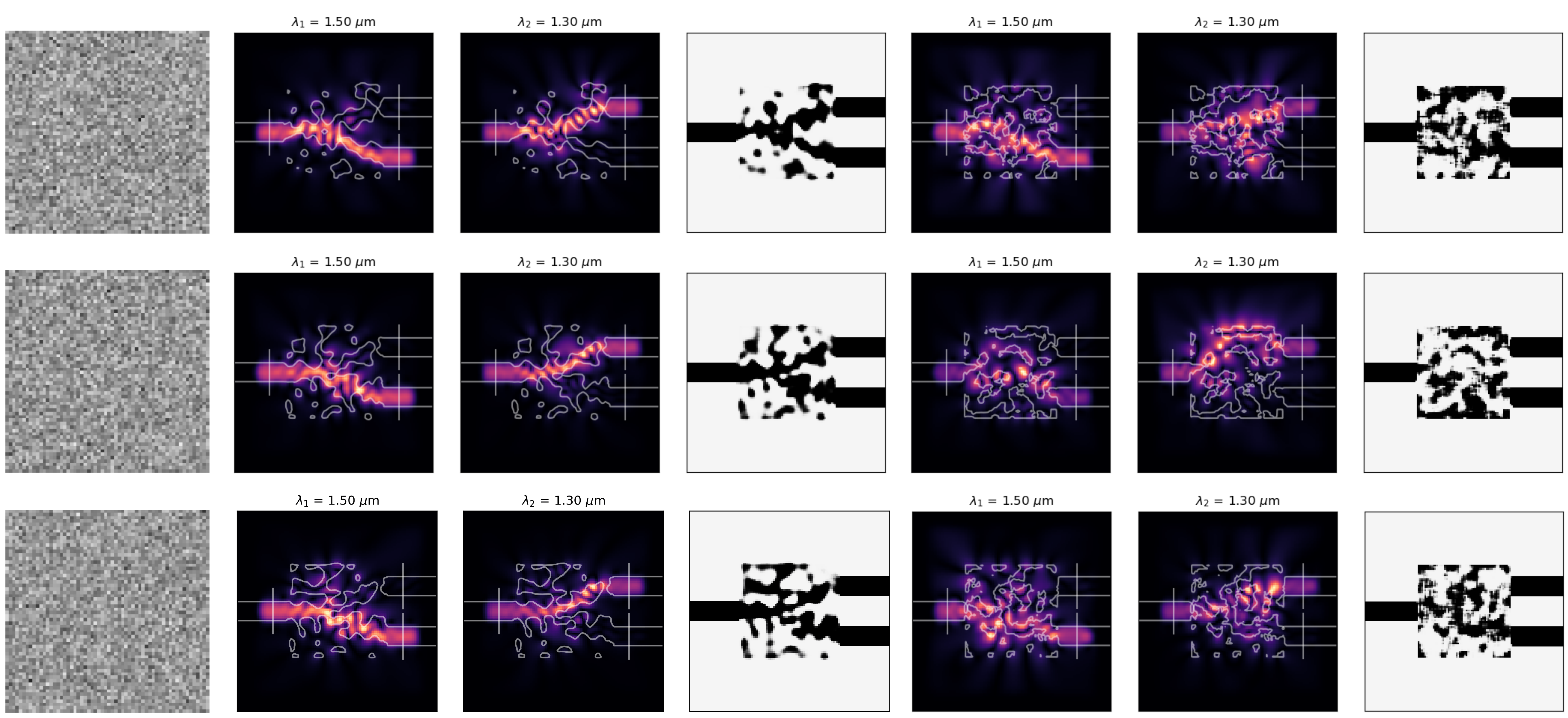}
  \caption{A sample collection of optimized permittivity distribution and the corresponding electric field intensity given initial Gaussian noise $N(0,0.1)$.}
\label{fig:Gaussian01}
\end{figure}

\begin{figure}[!ht]    
    \centering
\includegraphics[width=1\textwidth]{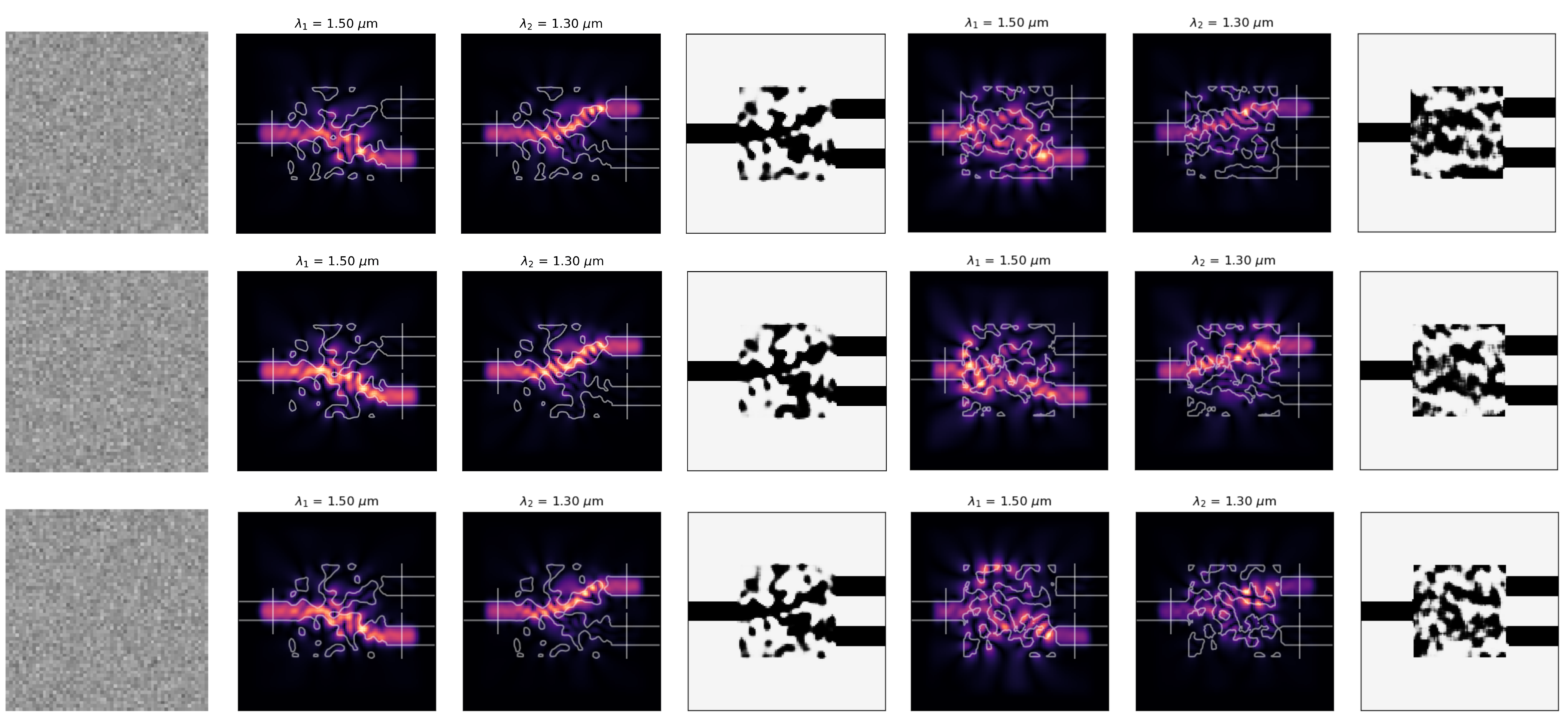}
  \caption{A sample collection of optimized permittivity distribution and the corresponding electric field intensity given initial Gaussian noise $N(0,0.05)$. }
\label{fig:Gaussian005}
\end{figure}

\begin{figure}[!ht]    
    \centering
\includegraphics[width=1\textwidth]{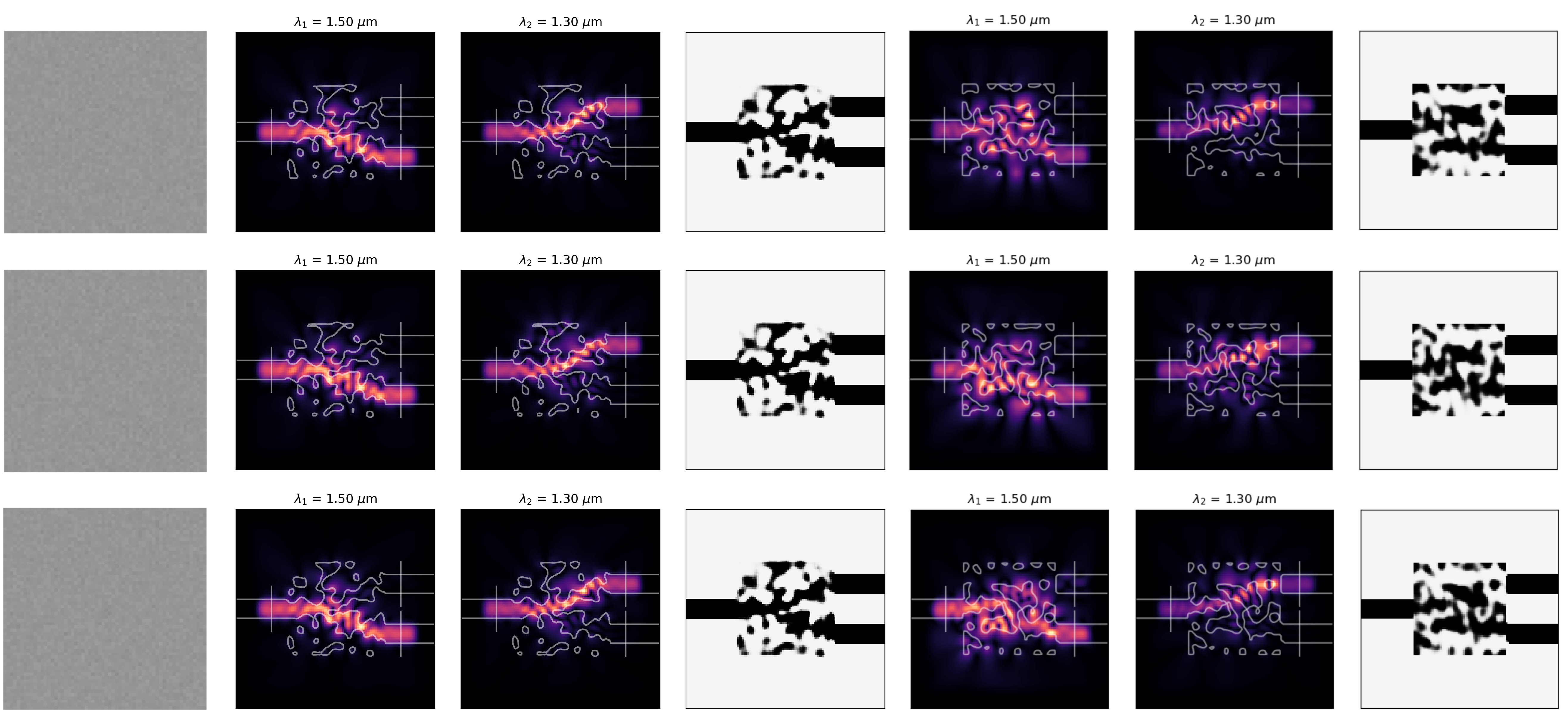}
  \caption{A sample collection of optimized permittivity distribution and the corresponding electric field intensity given initial Gaussian noise $N(0,0.01)$. }
\label{fig:Gaussian001}
\end{figure}

Figure~\ref{fig:Gaussian01} - Figure~\ref{fig:Gaussian001} show a sample collection of optimized design with the electric field intensity given Gaussian noise $N(0,0.1)$, $N(0,0.05)$ and $N(0,0.01)$ respectively. For a specific noise level, three selected samples are provided, and each of samples corresponds to a random {\color{black}initial guess} (first column), {\color{black} DGS-based nonlocal optimized design (column 4) with the electric field intensity at 1500 nm wavelength (column 2) and 1300 nm wavelength (column 3), as well as the local gradient-based design (column 7) with the electric field intensity at 1500 nm wavelength (column 5) and 1300 nm wavelength (column 6).} In particular, when the noise level is relatively high, as shown in Figure~\ref{fig:Gaussian01}, the optimized structure using the DGS gradient shows a clear binary distribution with an overall consistent pattern, while some small features and differences exist. The corresponding electric field shows a similar confined and clear transmission path. However, the optimized design using the local gradient method greatly varies with the random initialization. It is difficult to see a clear path as several noise features are embedded into the structure. This results in the fact that the electric field spreads across the entire device, suggesting that multi-path interference contributes to unexpected device performance. This is because the local gradient method is limited to escape the local minima that strongly depend on the {\color{black}initial guess}. Although noise level is decreased from $N(0,0.1)$ to $N(0,0.05)$, as shown in Figure~\ref{fig:Gaussian005}, the local gradient method is still affected by the initial random noise. As a result, the optimized structure includes a few small features with noise and the overall structure is unstable and unclear. On the contrary, the nonlocal optimization method is well-suited to handle the noise using a Gaussian smoothing operator with a large radius and thus achieves a robust and binarized design. When a small noise is imposed into the initialization, three samples from the nonlocal optimization method with the DGS gradient in Figure~\ref{fig:Gaussian001} are almost identical and converged to the optimized design as similar as the result using homogeneous initialization without any randomness. For the local gradient method, the noised feature is mitigated from the optimized structure but the resulting electric field intensity still underperforms the nonlocal method, specifically at the 1500 nm.

\subsubsection{Inverse design optimization with volume constraint}
\begin{figure}[!ht]    
    \centering
\includegraphics[width=0.35\textwidth]{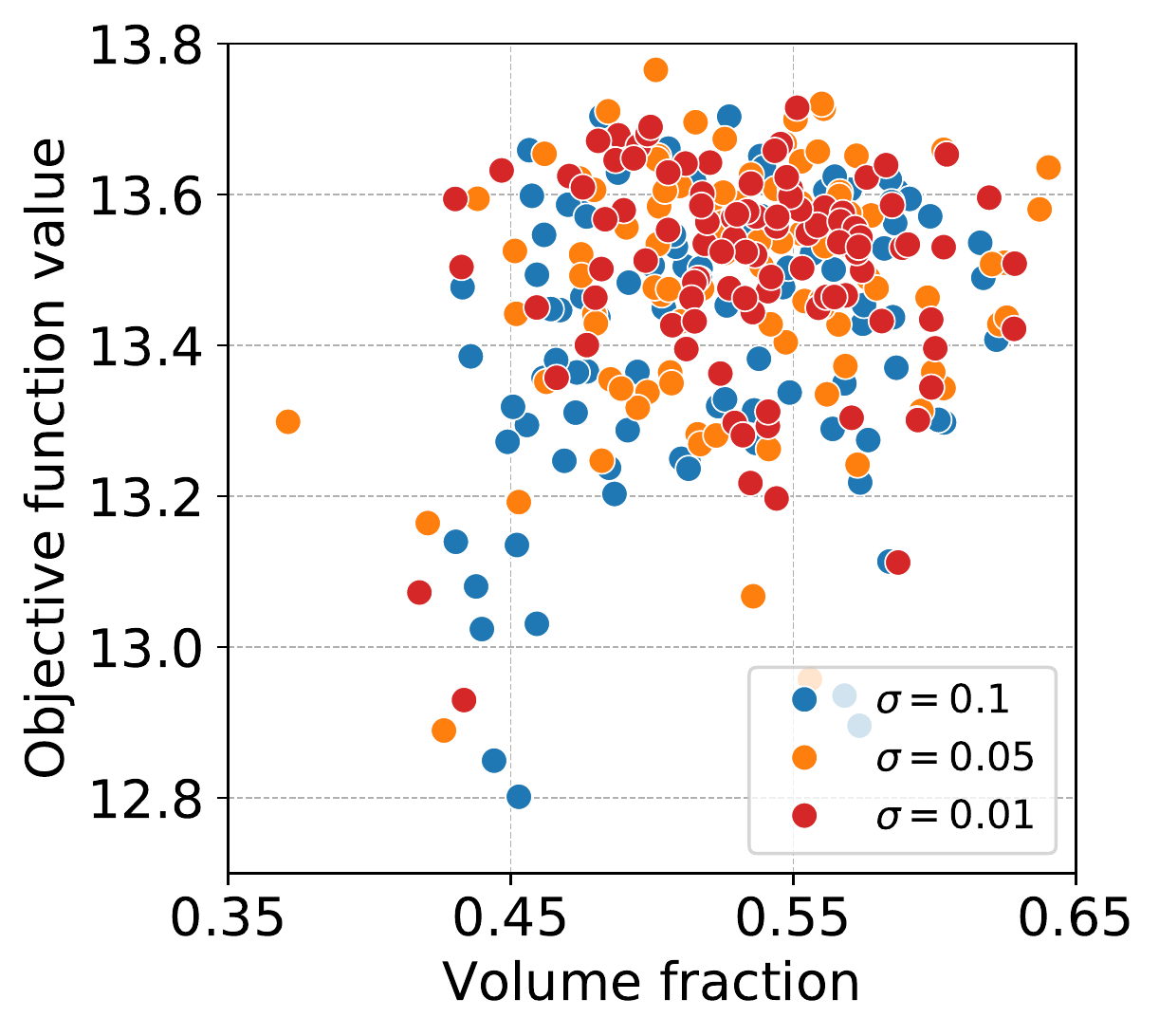}
  \caption{Relationship between volume fraction and optimized objective value under random initialization. There is no clear correlation between the used amount of materials and the optimized performance. }
\label{fig:vol}
\end{figure}

The optimized design discussed above is achieved by optimization without any constraint of materials usage. To investigate the effect of material usage on device performance, we conduct a study to investigate the relationship between volume fraction of material usage and objective function values. {\color{black} The general volume constraint in Eq.~\eqref{eq:vol} can be added to Eq.~\eqref{eq:obj}, which is given by
\begin{equation}
    h(\mathbf{x}) = \frac{V(\mathbf{x})}{V_0} - \gamma  = \frac{\sum_{i=1}^{N_p}\mathbf{x}}{N_p} - \gamma \le 0, \quad \mathbf{x} \in [0,1] \label{eq:vol2} 
\end{equation}
where $N_p$ is the total number of pixels, $V$ is the volume of material usage, $V_0$ is the original homogeneous distribution where all the computational design variables $\mathbf{x}=1$ and $\gamma$ is the specific constant that is the volume constraint fraction. Adding Eq.~\eqref{eq:vol2} to Eq.~\eqref{eq:obj}, the design problem is transferred from unconstrained optimization to constrained optimization problem and the materials usage can be controlled by assigning a specific value of $\gamma$.} As shown in Figure~\ref{fig:vol}, a total of 300 samples of three levels of Gaussian noise are used in this case but unfortunately, we didn't observe a strong correlation between the volume fraction and final objective values. Most cases of volume fraction $\gamma$ in Figure~\ref{fig:vol} concentrates in 0.45$-$0.65 and no cases are lower than 0.35. It naturally gives rise to an interesting question: is that possible to use fewer materials but achieve equivalently good performance? 

\begin{figure}[!ht]    
    \centering
\includegraphics[width=0.8\textwidth]{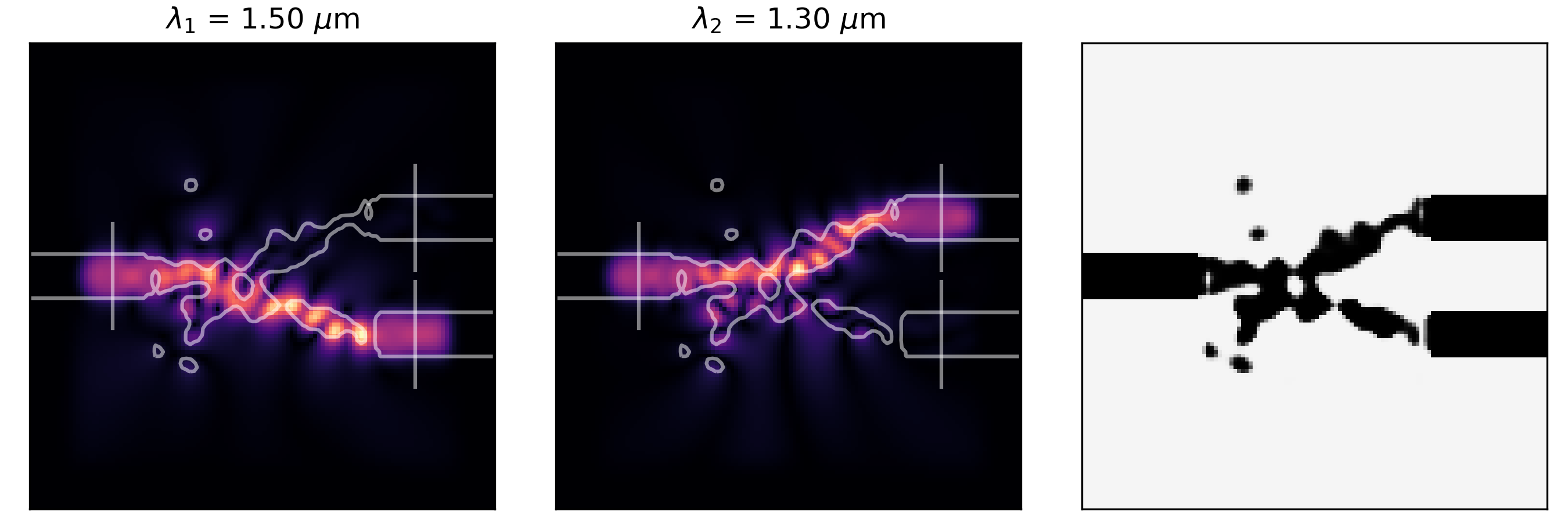}
  \caption{Optimized device with volume constraint $\gamma=0.3$. The electric field intensity of the optimized device at 1500 nm (left) and at 1300 nm (midde), as well as the optimized permittivity distribution (right) using DGS gradient with MMA optimizer}
\label{fig:vol_03_design}
\end{figure}

To answer this question, we reformulate the optimization by adding a volume constraint and solve this constrained optimization problem using the Method of Moving Asymptotes (MMA) \cite{svanberg1987method}. {\color{black} MMA is the state-of-the-art optimizer, which has been demonstrated to be versatile and well suited for wide range engineering design problems, particular in topology optimization. The basic of MMA aims at solving general nonlinear constrained optimization problem: 
\begin{equation}
\begin{aligned}
\min_{\mathbf{x}} &: \quad f_0(\mathbf{x}) + a_0z + \sum_{i=1}^m(c_iy_i + \frac{1}{2}d_i y_i^2)\\
s.t. &: \quad f_i(\mathbf{x})-a_i z -y_i \le 0,\quad i = 1,...,m \\
&: \quad \mathbf{x}\in X, \mathbf{y}\ge0, z\ge 0
\label{eq: mma1}
\end{aligned}
\end{equation}
Here, $X = \left\{x \in \mathbb{R}^n | x_j^{\min} \le x_j \le x_j^{\max}, j=1,...,n \right\}$, where $x_j^{\min}$ and $x_j^{\max}$ are given real numbers which satisfy $x_j^{\min} < x_j^{\max}$ for all $j, f_0, f_1, ..., f_m$ are given, continuously differentiable, real-valued functions on $X$, $a_0, a_i, c_i$ and $d_i$ are given real numbers which satisfy $a_0>0, a_i \ge 0, c_i \ge 0$ and $d_i \ge 0$ and $c_i + d_i >0$ for all $i$ and also $a_ic_i >a_0$ for all $i$ with $a_i>0$.  

MMA is a gradient-based method for solving Eq.~\eqref{eq: mma1} using the following steps. In each iteration, given the current point $(\mathbf{x}^{(k)}, \mathbf{y}^{(k)},z^{(k)})$, MMA generates an approximating subproblem, where the functions $f_i(\mathbf{x})$ are replaced by convex functions $\hat{f}_i^{(k)}(\mathbf{x})$. The approximating functions are determined by the {\em gradient information} at the current iteration point and moving asymptotes parameters which are updated in each iteration based on information from previous iteration points. The next iteration point $(\mathbf{x}^{(k+1)}, \mathbf{y}^{(k+1)},z^{(k+1)})$ is obtained by solving the subproblem, which is defined in \cite{svanberg1987method}.} However, MMA is limited to seek optima using local gradients information, either via adjoint method or finite difference. We address this challenge by inserting the DGS gradient into the MMA optimizer so that we can exploit the nonlocal exploration of the DGS operator to search for a better design. Figure~\ref{fig:vol_03_design} shows the optimized design with volume constraint $\gamma=0.3$ that means only 30\% of materials in the design domain are used. We easily observe a clear and confined light transmission path at both 1500 nm and 1300 nm. The optimized permittivity distribution shows the fundamental splitter-like feature and eliminates a lot of unnecessary material in the device compared with the design without volume constraint. 

\begin{figure}[!ht]    
    \centering
\subfigure[]{\includegraphics[width=0.31\textwidth]{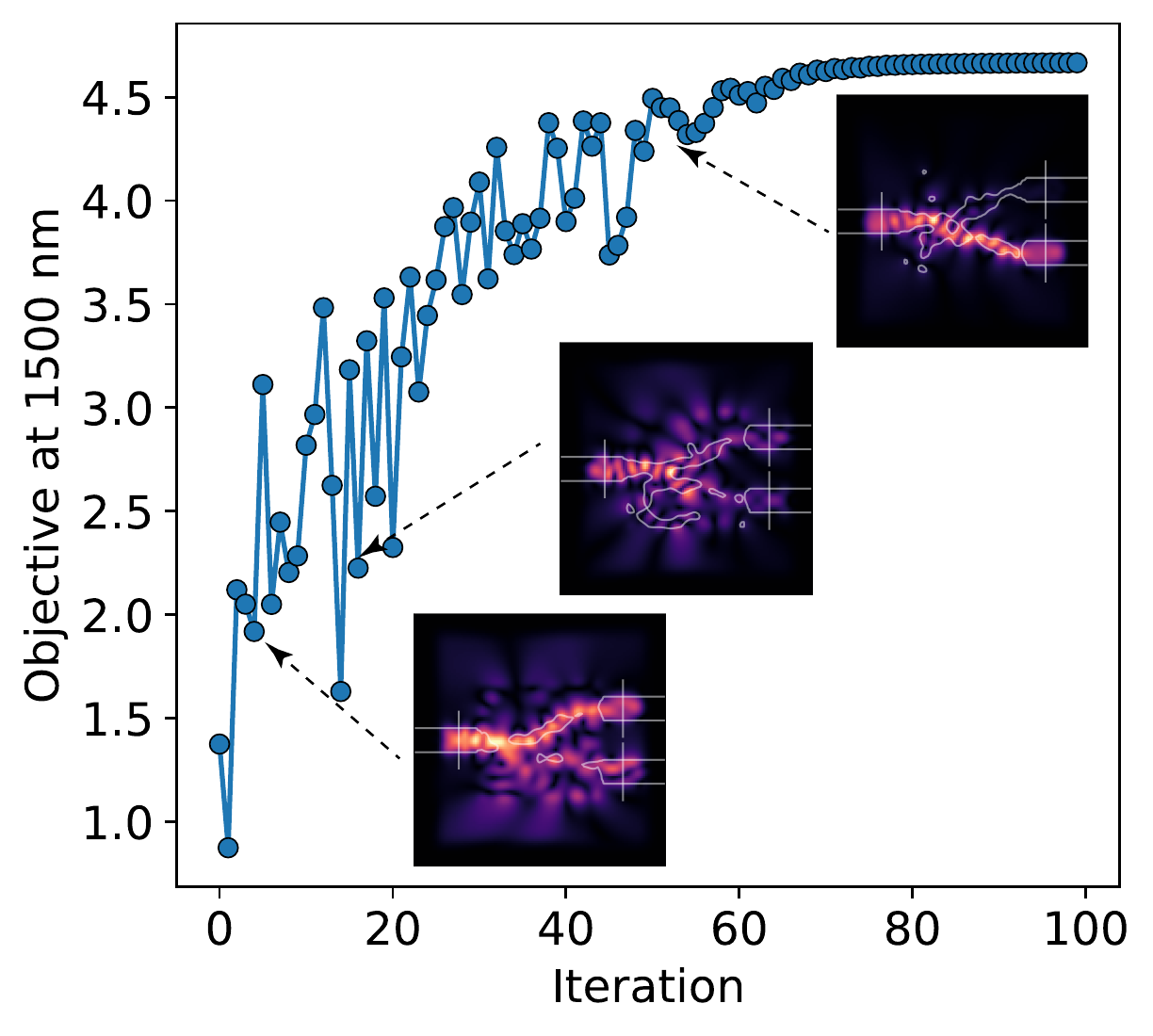}}
\subfigure[]{\includegraphics[width=0.31\textwidth]{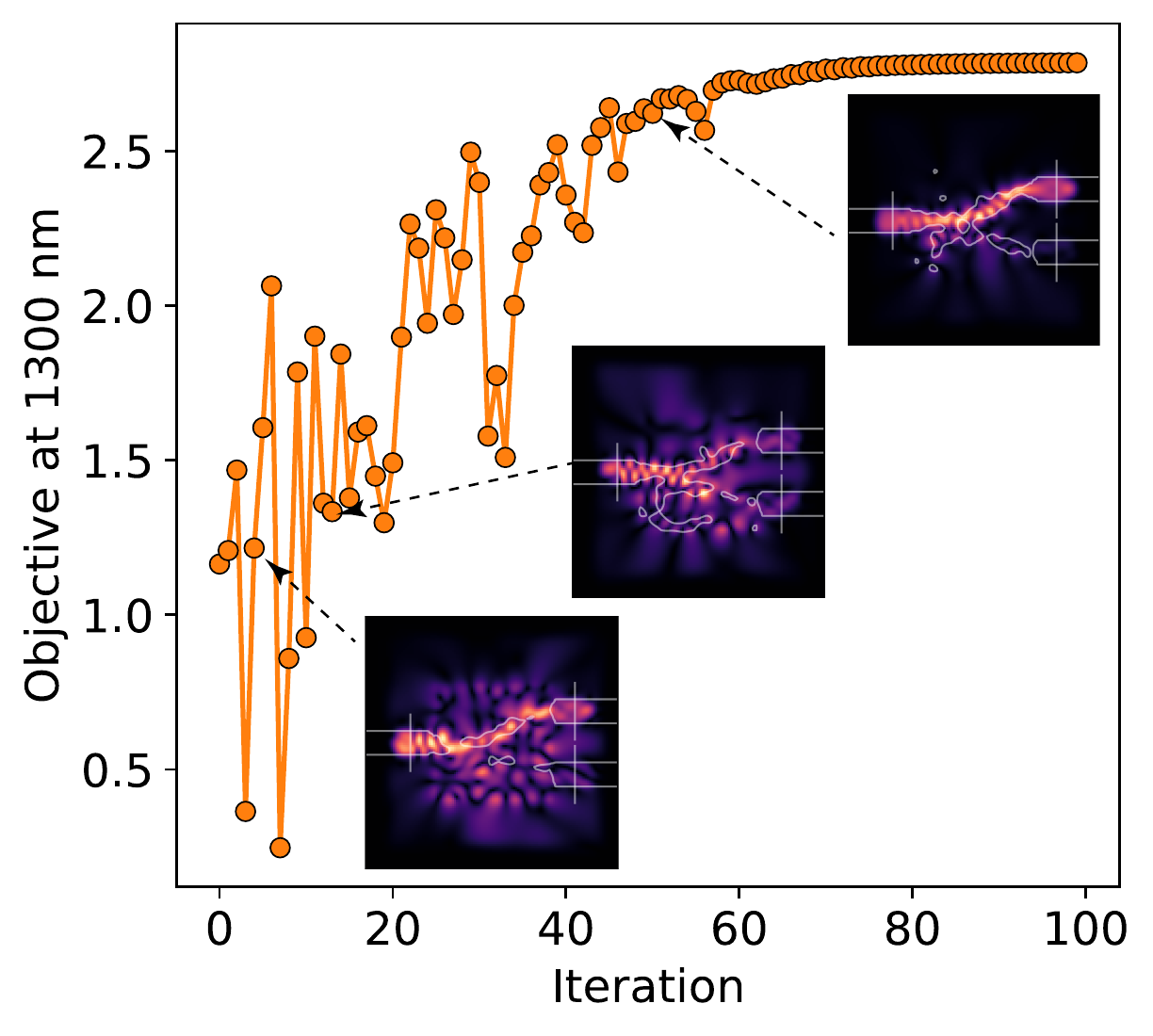}}
\subfigure[]{\includegraphics[width=0.36\textwidth]{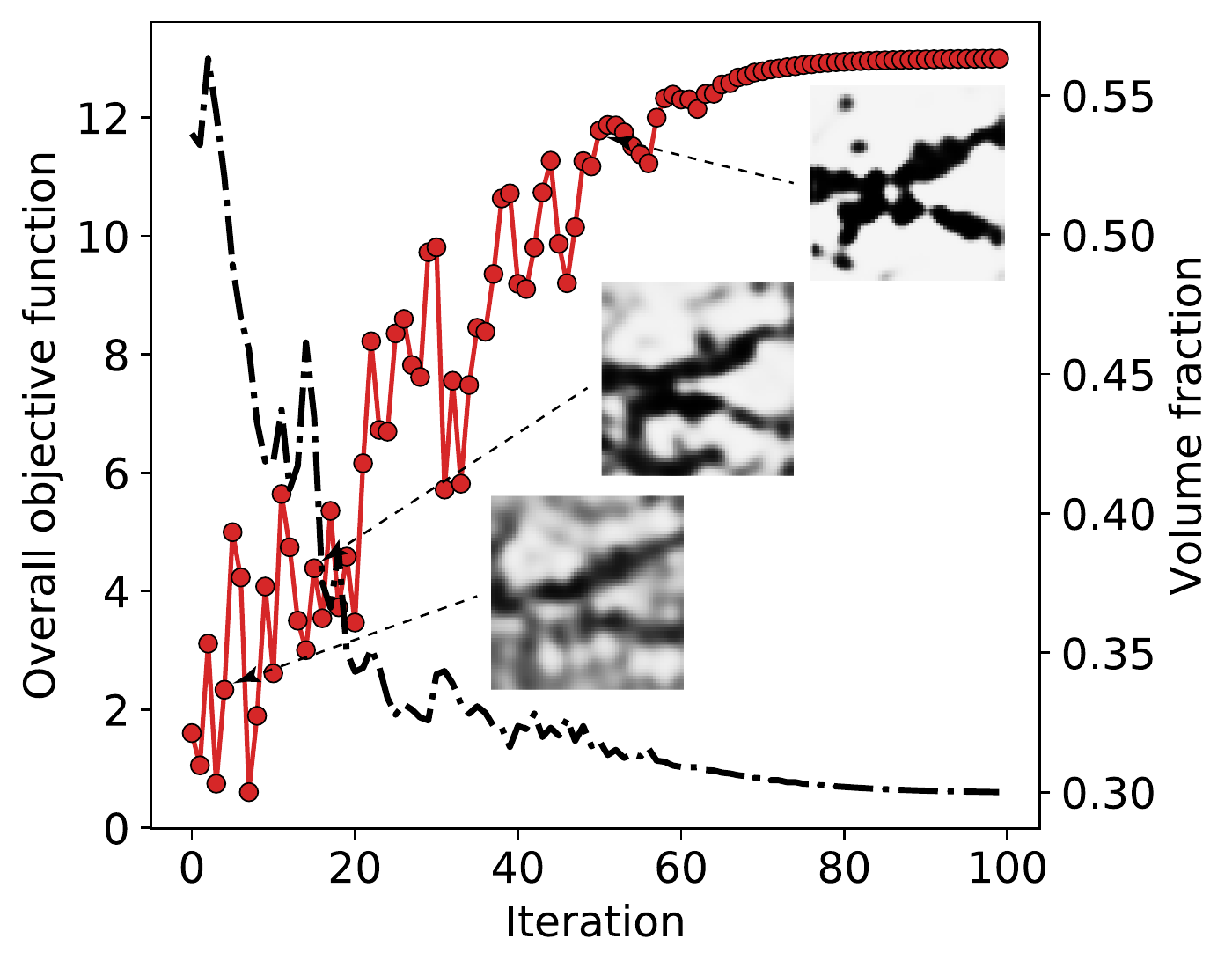}}
  \caption{Iteration history of constrained optimization with $\gamma=0.3$ and the corresponding electrical field intensity and permittivity distribution. (a) Objective at 1500 nm, (b) objective at 1300 nm and (c) overall objective and volume fraction iteration.}
\label{fig:vol_03_iter}
\end{figure}

Figure~\ref{fig:vol_03_iter} shows the iteration history of constrained optimization with $\gamma=0.3$.  To illustrate the optimized process in detail, Figure~\ref{fig:vol_03_iter} (a) shows the iterative change of electric field intensity at 1500 nm and the objective value that corresponds the term $\exp \left[ \log(\mathbf{c}^{\dagger }\mathbf{E}_1) - \log(\mathbf{c}^{\dagger }\mathbf{E}_{z_1}) \right]$ in objective function (see Eq.~\eqref{eq:obj}) . Similarly, the iteration history of objective term $\exp \left[ \log(\mathbf{c}^{\dagger }\mathbf{E}_2) - \log(\mathbf{c}^{\dagger }\mathbf{E}_{z_2}) \right]$ at 1300 nm is shown in Figure~\ref{fig:vol_03_iter} (b). Note that, under the volume fraction constraint, the iteration curves show some oscillations initially but quickly converge at the 60th iteration. The designed device successfully separate both signals at the assigned port. Figure~\ref{fig:vol_03_iter} (c) shows the iteration history of overall objective (red color), the decay of volume fraction (black color) and the changes of permittivity distribution. The constrained optimization using DGS gradient integrating with MMA optimizer achieves a nearly same high performance, $f^{0.3}_{\textup{DGS}} = 13.09$ as the optimization without volume constraint. But the amount of material usage is significantly reduced from 0.474 (the case in Figure~\ref{fig:dgs_adam}(a)) to 0.3 and we therefore save 36.7\% material usage. 

\begin{figure}[!ht]    
    \centering
\includegraphics[width=0.8\textwidth]{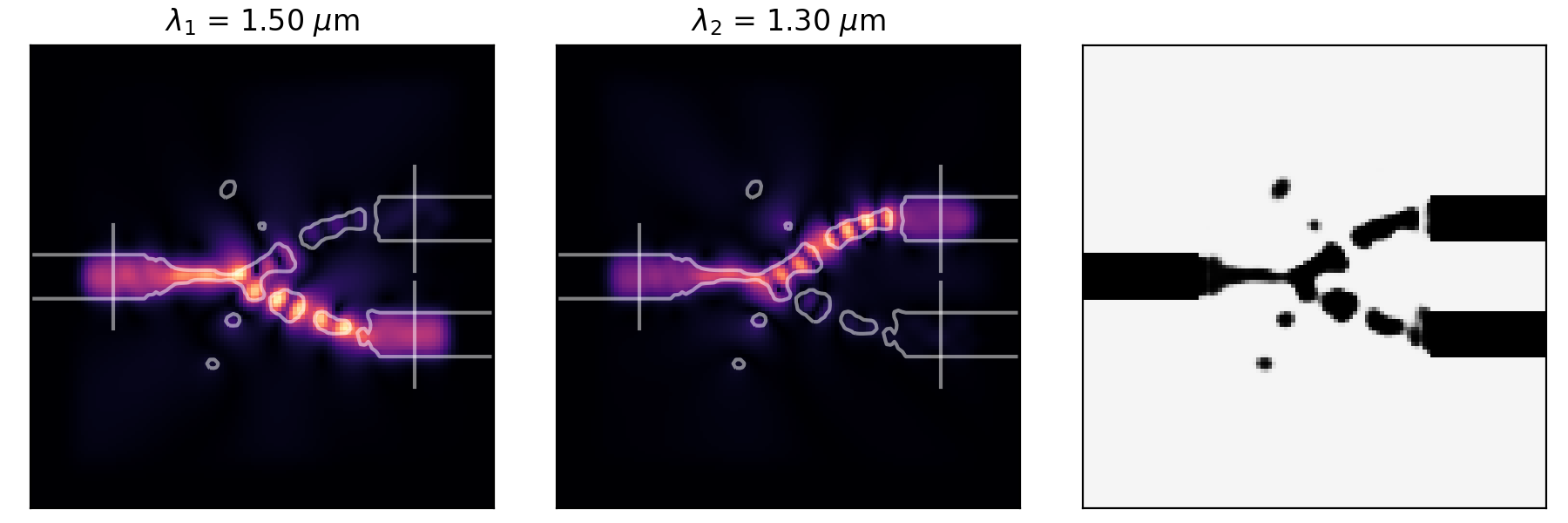}
  \caption{Optimized device with volume constraint $\gamma=0.2$. The electric field intensity of the optimized device at 1500 nm (left) and at 1300 nm (midde), as well as the optimized permittivity distribution (right) using DGS gradient with MMA optimizer}
\label{fig:vol_02_design}
\end{figure}

We further reduce the volume fraction from 0.3 to 0.2 to exploit the maximizing capability of the nonlocal method using the DGS gradient integrating with the MMA optimizer. Figure~\ref{fig:vol_02_design} shows the electric field intensity based on the optimized permittivity distribution given $\gamma=0.2$. The optimized design still retains the principle splitter-like features that connect the input port and two output ports, even though a very limited amount of material is used. As similar to the optimized design with $\gamma=0.3$, few small spots are disconnected from the main structure and probably play a limited role in the effective transmission of the input source. This special feature, observed by several previous studies \cite{Piggott2015,Su2020,Su2018,hughes2018adjoint} is probably due to the issue of local minima. Although the DGS operator enables nonlocal exploration to facilitate the global search, the optimized devices may be trapped into one of the local minima. 

Figure~\ref{fig:vol_03_iter} shows the iterative process of constrained optimization. It is noted that the iteration history at 1500 nm (Figure~\ref{fig:vol_03_iter} (a)) and 1300 nm (Figure~\ref{fig:vol_03_iter} (b)) shows a relatively large oscillation, specifically in the initial stage. This probably results from the fast decay of volume fraction in the initial period, as shown in Figure~\ref{fig:vol_03_iter} (c). After 60 iterations, the objective values tend to converge and the volume fraction also approaches to the constrained value, $\gamma=0.2$. This case with a smaller volume fraction achieves a final objective value $f^{0.2}_{\textup{DGS}} =12.70$ that is slightly lower ($\sim3.71\%$) than the case of $\gamma=0.3\%$ but saves $33.3\%$ material usage. 

\begin{figure}[!ht]    
    \centering
\subfigure[]{\includegraphics[width=0.31\textwidth]{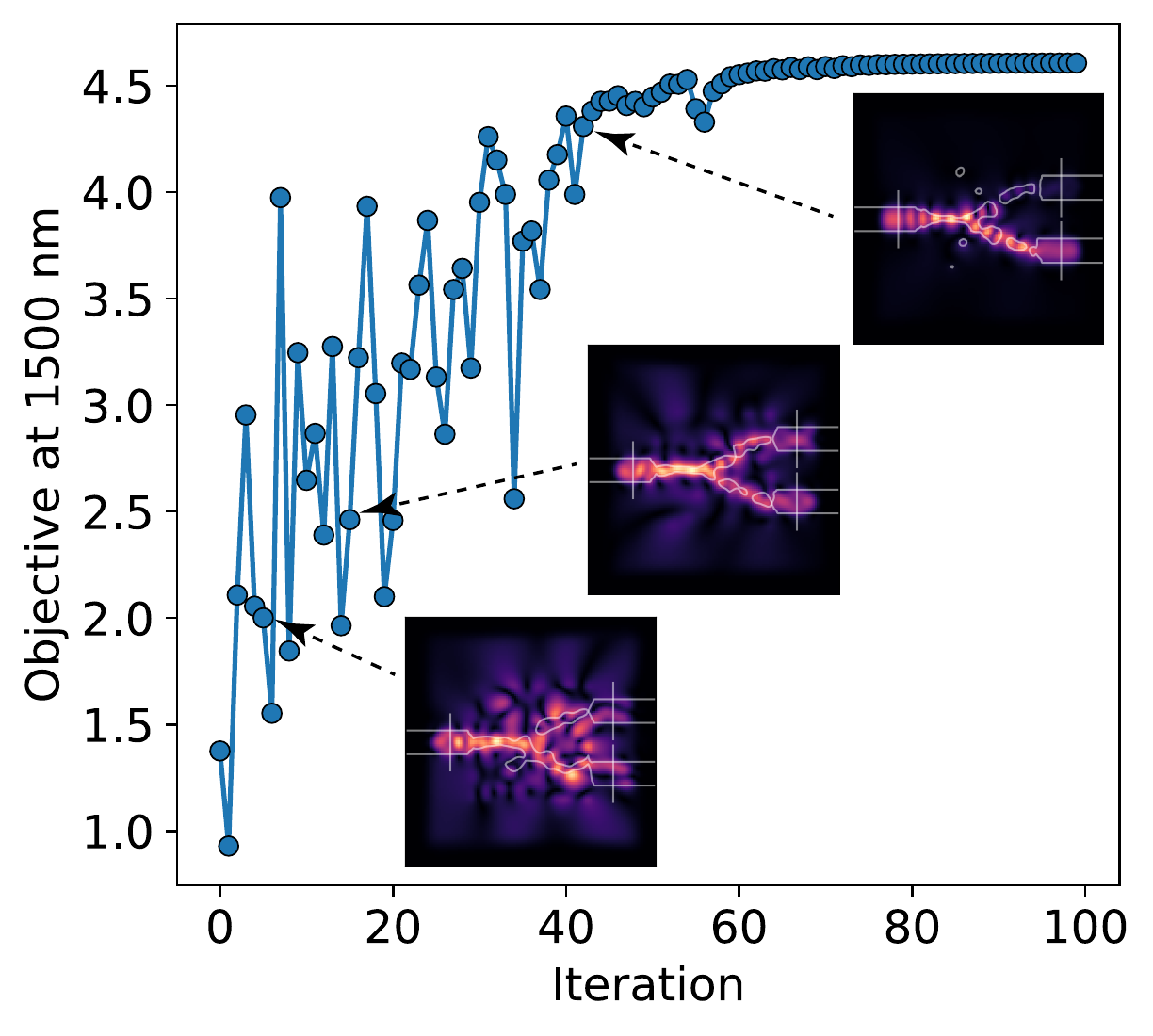}}
\subfigure[]{\includegraphics[width=0.31\textwidth]{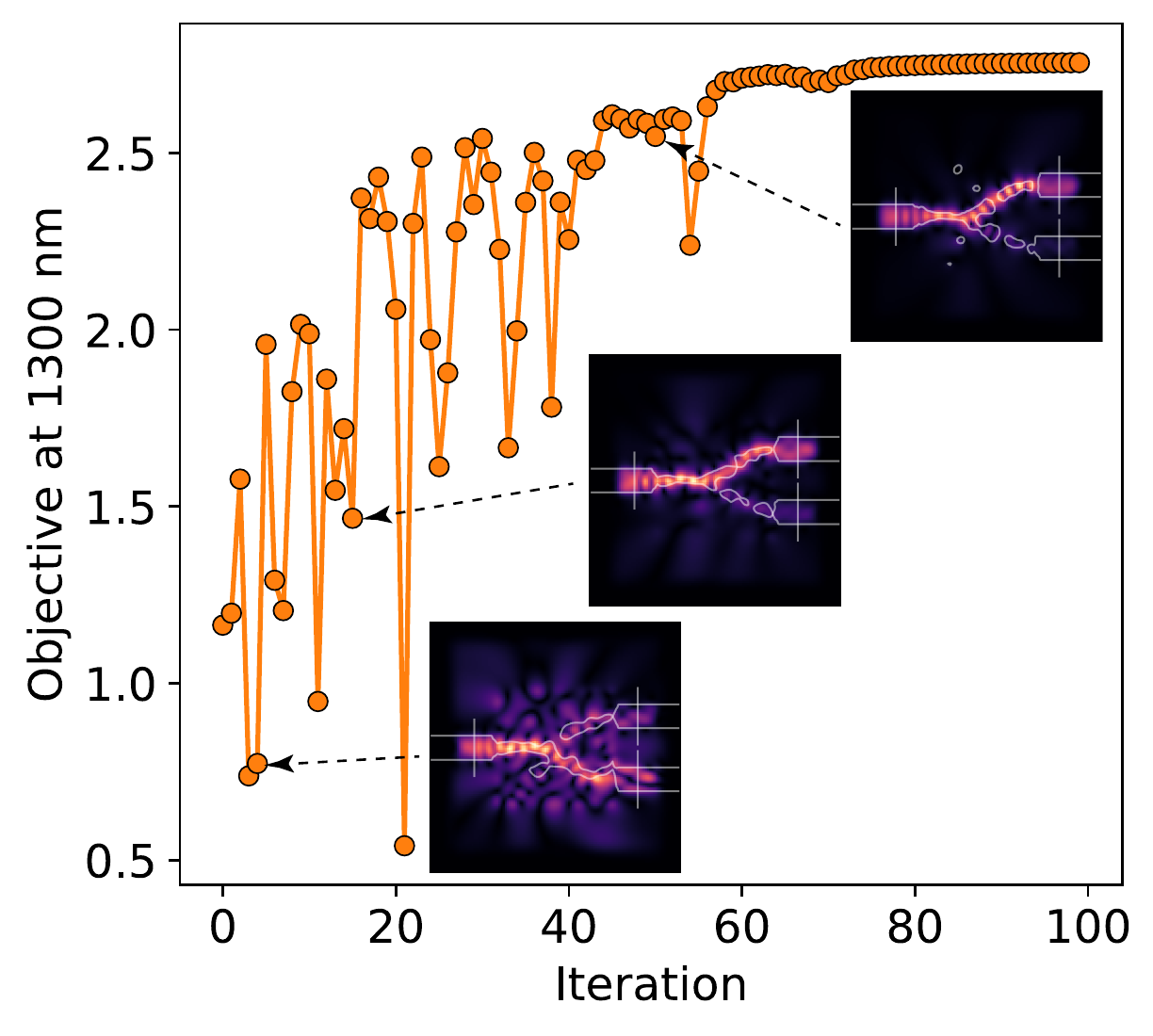}}
\subfigure[]{\includegraphics[width=0.36\textwidth]{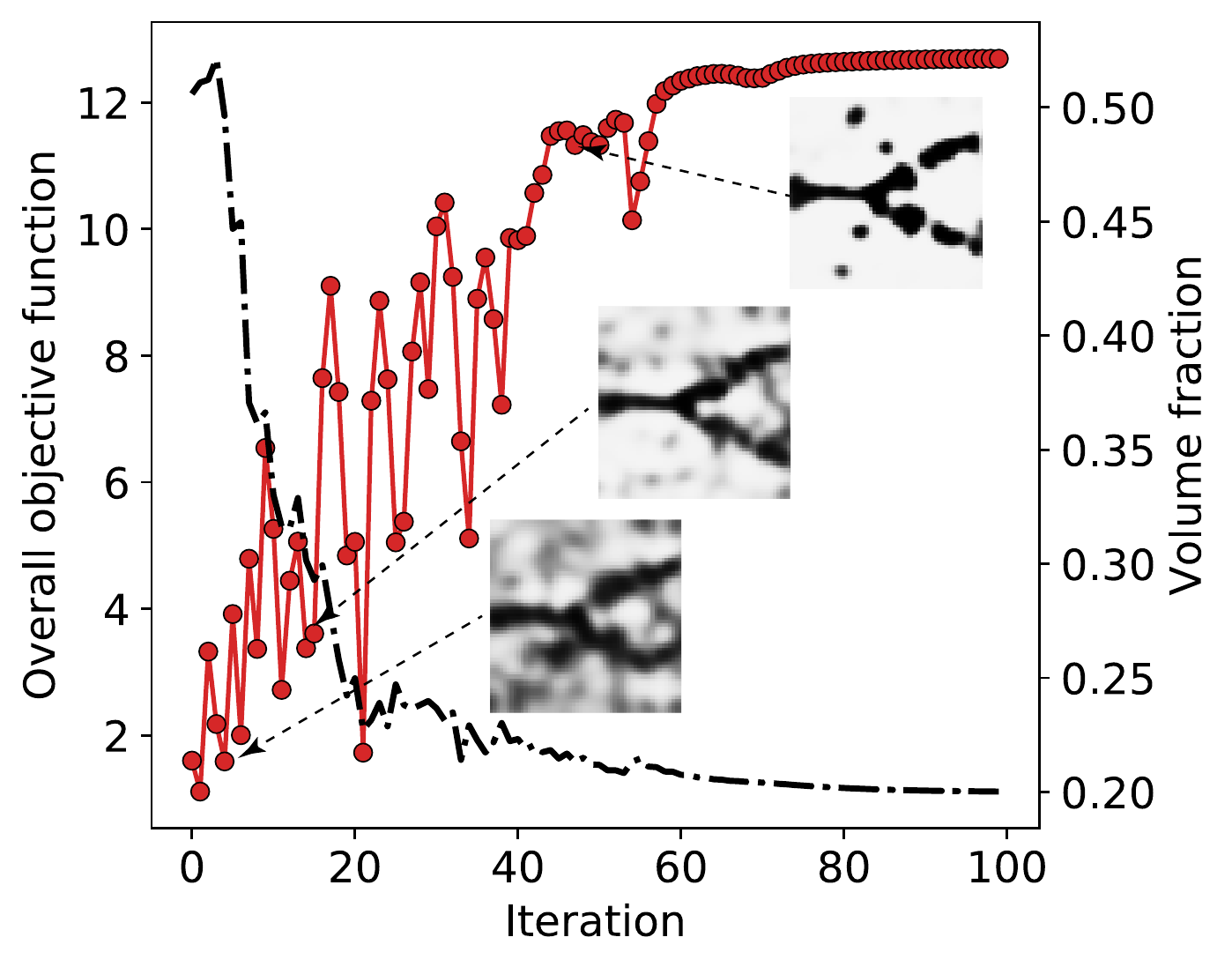}}
  \caption{Iteration history of constrained optimization with $\gamma=0.2$ and the corresponding electrical field intensity and permittivity distribution. (a) Objective at 1500 nm, (b) objective at 1300 nm and (c) overall objective and volume fraction iteration.}
\label{fig:vol_02_iter}
\end{figure}

\begin{figure}[!ht]    
    \centering
\subfigure[]{\includegraphics[width=0.31\textwidth]{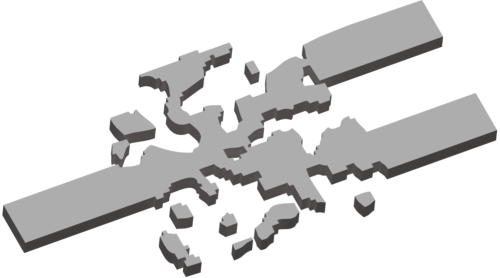}}
\subfigure[]{\includegraphics[width=0.31\textwidth]{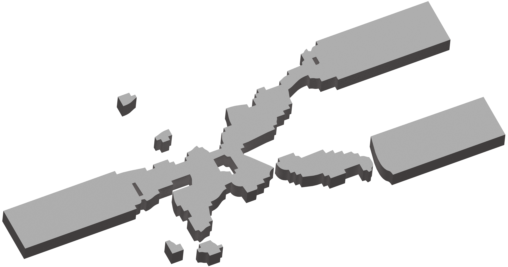}}
\subfigure[]{\includegraphics[width=0.31\textwidth]{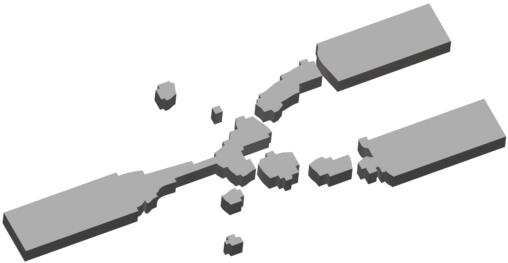}}
  \caption{A three-dimensional rendering of the optimized design. Silicon is shown in grey and light enters the optimized device from the input waveguide on the left-hand side (port 1) and exits via one of the two output waveguides (port 2 and port 3) on the right side. There are three optimized devices with different volume fraction (a) $\gamma = 0.474$, (b) $\gamma = 0.3$ and (c) $\gamma = 0.2$.}
\label{fig:3D}
\end{figure}

The final designed devices with three different amounts of material usage are shown in Figure~\ref{fig:3D}. From the angled view, the vertical sidewalls are clearly visible and the permittivity distribution along the vertical direction are all same. This is easy for fabricating using electron-beam (top-down) lithography followed by etching the 220-nm-thick layer of an SOI substrate, leaving the structure with an air cladding. Compared with the classical design approaches, the proposed inverse design framework integrating the DGS gradient with MMA optimizer provides higher final performance with less material usage for real fabrication in practice. 

{\color{black}
\section{Discussion and limitation}
This section will address several concerns in the practical implementation of the proposed method and the existing limitations with the current version of the DGS algorithm in computational inverse design.

\subsection{Hyperparameters choice}
Hyperparameters used in the proposed method play important roles in the final optimized performance. In the DGS gradient operator, the most sensitive hyperparameter is the smoothing radius of $\sigma_r$. If $\sigma_r$ is too small, the loss function will be insufficiently smoothed, such that the optimizer may be trapped in a local minimum. In contrast, if $\sigma_r$ is too big, the loss function is overly smoothed, such that the convergence will become much slower. Based on our hyperparameter tuning experience, our suggestion is to set up the smoothing radius $\sigma_r$ according to the search domain. In the photonic design problem, the computational domain is $\mathbf{x}$ and its design domain is $[0,1]$. Specifically, we adopted a quadratic decay schedule ($\alpha=2$) for smoothing radius $\sigma_r$ and learning rate $\ell_r$. The initial value of smoothing radius $\sigma_r$ is set to $20\%-25\%$ of the search domain width, and the ending value of smoothing radius $\sigma_r$ is set to $1\%-5\%$ of the search domain width. The learning rate of $\ell_r$ is the second important parameter that affects the convergence performance. Similarly, we determine the learning rate $\ell_r$ according to the search domain and set the initial value to be $10\%$ and ending value to be $1\%$ of the search domain width. Although the Gauss-Hermite quadrature is the key to the overall superior performance, we consider that 5 GH points already have a high estimate accuracy and can be used for most of the problems. 

The hyperparameters in the parameterization scheme will also affect the final performance in terms of the fabrication constraints. Typically, the projection strength and convolution filtering radius depend on the computational design domain $\mathbf{x}$. A general guideline to choose the projection strength is that the initial value $\beta^{\textup{ini}}$ can be $20\%-25\%$ and the ending value $\beta^{\textup{end}}$ can be $1\%-5\%$ of the design domain width. The filtering radius $\alpha=2$ is typically used and the effect of the different radius is discussed, as shown in Fig.4. We realize that the hyperparameter tuning is a non-trivial task and how to adaptively adjust these parameters is still an open question. 

\subsection{Computational cost challenge}
Differing from the local gradient method, the computational cost of the DGS method linearly depends on the dimensionality of the design space. At each iteration, a relatively large number of function evaluations are required to estimate the nonlocal gradient using a smoothing strategy. Using a big learning rate, DGS can potentially achieve a faster convergence with fewer iterations but the total computational cost is still large. We realize this is a critical challenge, specifically given a small computational budget. This computational challenge can be mitigated by using parallel computing because all evaluations at each iteration are completely parallelizable as those in random sampling with very small communication costs. In this work, we implement the DGS method by using 44 core CPUs but it can be easily scaled to thousands of CPUs on Supercomputers, like OLCF Summit, which has 9,216 POWER9 22-core CPUs. Ideally, each function evaluation can be deployed to one CPU so that DGS achieves an equivalent cost to the local gradient at each iteration. Due to its ideal scalability, DGS is well-suite to the applications where the simulation or modeling can make use of high performance computing (HPC) if large computational resources are available. In the next version of the DGS method, we will incorporate dimension reduction techniques, e.g., active subspace \cite{constantine2015active}, and nonlinear level-set method \cite{zhang2019learning}, to reduce the dimensionality requiring directional smoothing, which can alleviate the dependence on computing resources and finally overcome the limitation of the relatively large computational cost. 

\subsection{Optimization performance and sub-optimal solution}

Although the computational cost is relatively large, the DGS shows superior capabilities to explore the better solution, which is more critical to address the design challenges by optimization. The local gradient method performs efficiently and converges fast, but it often traps to a local minimum that underperforms the DGS method. In addition, DGS shows better robustness than the local gradient method, which is sensitive to the choice of initial guess and the final optimized design varies largely. For the derivative-free global optimization algorithms, e.g., GA or PSO, they are also very computationally intensive and frequently fail to find an optimal solution in such high dimensional space. To some extent, these evolutionary algorithms show a similar performance as the ``random search" method. Like Bayesian optimization (BO), it is intractable to handle more than thousands of dimensional problems due to the limitation in Gaussian process modeling. The constraints in the BO framework and stochastic optimization such as SGD and Adam is still a critical issue. We realize that there is a trade-off between computational cost and final optimized performance, but we consider it is more important to first search a better solution rather than a worse solution but saving time. 

The proposed DGS method has advantages to capture the global structure in the loss landscape. However, if the loss function without global structures, for example, the Schwefel function, as shown in \cite{2020arXiv200203001Z}, the DGS method cannot find the global minimum. This could happen in real-world inverse design problems and other engineering applications. Even though our method outperforms the local gradient method in solving the nanophotonic design problem, we cannot verify that the final design optimized by our method is globally optimal. 
}

\section{Conclusion}
This work focuses on the development of a nonlocal optimization method for computational inverse design in nanophotonics. A novel DGS gradient operator is introduced to improve the nonlocal exploration capability required for escaping from local minima in the high-dimensional non-convex landscapes. The DGS gradient operator is achieved by conducting 1D nonlocal explorations along with $d$ orthogonal directions in $\mathbb{R}^d$, each of which defines a nonlocal directional derivative as a 1D integral. Instead of Monte Carlo (MC) sampling, a deterministic Gauss-Hermite (GH) quadrature is used to estimate each of 1D integrals in $d$-dimension to achieve high accuracy. Compared with the local gradient method, the directional smoothing allows for a large smoothing radius to capture the global structure of loss landscapes. GH quadrature with error bound provides guarantees higher accuracy than random sampling, even though a large smoothing radius is used. 

The nonlocal optimization method using the DGS gradient has advantages in portability and flexibility so that it is naturally incorporated with parameterization, physics simulation, and objective formulation to build up an effective optimization workflow for inverse design. Within the nonlocal optimization scheme, an adaptive smoothing radius and learning rate with quadratic decay is proposed to accelerate the convergence and improve the robustness by reducing the dependence of optimized design on random initialization. To make the optimized design easy to fabrication, a dynamic growth mechanism is imposed on the projection strength in parameterization to achieve a clear material layout. Moreover, we investigate the effect of material usage on optimized performance by integrating the DGS gradient with MMA optimizer to conduct a nonlocal constrained optimization with adding volume constraint into optimization formulation. 

The proposed method is demonstrated on a wavelength demultiplexer design problem that aims to split 1500 nm and 1300 nm signals from an input waveguide into two output waveguides. The results show that the proposed inverse design framework using the nonlocal optimization method achieves an improvement of approximate 10\% performance with faster convergence compared with the classical local gradient-based approaches. Given different levels of the random {\color{black}initial guess}, the DGS-based nonlocal method presents a smaller variation and higher robustness on the final optimized design. Optimization with volume constraint demonstrates the final optimized device can maintain the high performance as similar to the optimized design without volume concern but achieves a significant reduction (36.7\% for $\gamma=0.3$ and 57.8\% for $\gamma=0.2$) of material usage.  

The future work may potentially explore the effect of different types of random {\color{black}initial guess} on the local minima. The current study only addresses the Gaussian noise with different standard deviation to the {\color{black}initial guess}. It would be interesting to study the other types of random noise, for example, Perlin noise and Gabor noise as suggested in \cite{Su2020}. Besides, the optimized devices are composed of a small number of distinct spot materials, which may play an uncertain role in light transmission and separation. We plan to make a further investigation on the reason for these small features and explore their effect on the performance of electric field intensity, specifically when volume fraction is relatively low.


\section{Data availability statement}
The data of this study will be made available on request.

\section{Acknowledgements}
This work was supported by the U.S. Department of Energy, Office of Science, Office of Advanced Scientific Computing Research, Applied Mathematics program under contract ERKJ352 and ERKJ369; and by the Artificial Intelligence Initiative at the Oak Ridge National Laboratory (ORNL). ORNL is operated by UT-Battelle, LLC., for the U.S. Department of Energy under Contract DEAC05-00OR22725.

 \bibliographystyle{unsrt} 
\bibliography{sample.bib,reference.bib}


\end{spacing}
\end{document}